\documentclass[a4paper,11pt]{article}
\pdfoutput=1 

\usepackage{jcappub} 

\usepackage[T1]{fontenc} 
\usepackage[latin9]{inputenc}
\usepackage{babel}

\title{\boldmath Spherical collapse of non-top-hat profiles in the presence of dark
energy with arbitrary sound speed}

\author[a,b]{R. C. Batista}
\author[c]{H. P. de Oliveira}
\author[b]{L. R. W. Abramo}

\affiliation[a]{Escola de Ciências e Tecnologia, Universidade Federal do Rio Grande do
Norte, Campus Universitário Lagoa Nova, 59078-970, Natal, RN, Brazil.}
\affiliation[b]{Instituto de Física, Universidade de São Paulo, Rua do Matão 1371, 05508-030, São Paulo, SP, Brazil.}
\affiliation[c]{Instituto de Física A. D. Tavares,
Universidade do Estato do Rio de Janeiro, Rua São Francisco Xavier 524, 20550-013, Rio de Janeiro, RJ, Brazil.}

\emailAdd{rbatista@ect.ufrn.br}

\abstract{
We study the spherical collapse of non-top-hat matter fluctuations
in the presence of dark energy with arbitrary sound speed. The model
is described by a system of partial differential equations solved
using a pseudo-spectral method with collocation points. This method
can reproduce the known analytical solutions in the linear regime
with an accuracy better than $10^{-6}\%$ and better than $10^{-2}\%$
for the virialization threshold given by the usual spherical collapse model.
We show the impact of nonlinear dark energy fluctuations on matter
profiles, matter peculiar velocity and gravitational potential. We also show that phantom dark
energy models with low sound speed can develop a pathological behaviour
around matter halos, namely negative energy density. The dependence
of the virialization threshold density for collapse on the dark energy
sound speed is also computed, confirming and extending previous results in the limit
for homogeneous and clustering dark energy.}

\begin{document}
\maketitle
\flushbottom

\section{Introduction}

The Spherical Collapse (SC) model, as proposed by Gunn and Gott \cite{Gunn:1972sv},
describes the nonlinear evolution of pressureless matter perturbations
in Einstein-de-Sitter Universe (EdS). This model can be used to determine
the critical density of collapse, $\delta_{c}$, which can be used
in Press-Schechter or Sheth-Tormen \cite{Press:1973iz,Sheth:1999su}
halo mass functions to compute the abundance of Dark Matter (DM) halos
in the universe. However, the expansion of the universe is accelerating,
which indicates that, assuming General Relativity correctly describes
gravitational interactions on large scales, the universe is composed
of roughly $70\%$ Dark Energy (DE) and $30\%$ of matter (DM plus
barions) today. Even before the discovery of the accelerated expasion,
the SC model has been generalized to include the Cosmological Constant,
$\Lambda$, \cite{Lahav:1991wc,1992ApJ...386L..33L,Eke:1996ds,Kitayama1996}.
Later, homogeneous DE models described as a perfect fluid were studied,
e.g., \cite{Kitayama1996,Wang:1998gt,Weinberg2003,Percival:2005vm}.
In these scenarios, DE induces a small (at most $1\%$) decay of $\delta_{c}$
at low-$z$ in comparison to the standard EdS value ($\delta_{c}\simeq1.686$). 

Recently, observational data has indicated that the value of the Hubble
constant predicted by the $\Lambda$CDM model is in tension with local
astrophysical measurements. There also exists a less significant tension
related to the normalization of matter perturbations, expressed in
terms of the $S_{8}=\sigma_{8}\left(\Omega_{m}^{0}/0.3\right)^{1/2}$
parameter. See \cite{Abdalla:2022yfr} for a discussion and several
proposals to solve these issues. If the accelerated expansion is not
caused by $\Lambda$, DE necessarily has fluctuations, which might
be important for the evolution of matter fluctuations on small scales,
for a review, see \cite{Batista:2021uhb}. For a concrete recent study of how DE perturbations
can alleviate these tensions, see \cite{Cardona:2022pwm}.

Several papers have studied the SC model and halo abundances in the
presence of DE fluctuations, e.g., \cite{Mota:2004pa,Nunes:2004wn,Abramo2007,Mota:2007zn,Creminelli2010,Wintergerst2010a,Basse2011,Batista:2013oca,Pace2014a,Velten2014a,Batista2017,Heneka2018,Pace2022}.
The key parameter that determines the impact of DE fluctuations is
the sound speed $c_{s}=\sqrt{\delta p_{de}/\delta\rho_{de}}$. If
$c_{s}\simeq1$, DE fluctuations are much smaller than matter fluctuations
on small scales and essentially do not modify the growth of nonlinear
structures. The nonlinear evolution of Quintessence and Tachyon models
were studied in \cite{Rajvanshi2020}. For Quintessence, $c_{s}=1$,
and its perturbation remains very small even when matter perturbations
become nonlinear. As we will show, this also happens in the fluid
description implemented in this work. In Tachyon models, $c_{s}^{2}=-w$
and, since $w\simeq-1$ at late times, the sound speed is also near
the unity, erasing DE fluctuations on small scales.

On the order hand, if $c_{s}$ is sufficiently small, DE perturbation
can be of the same magnitude as matter fluctuations. If $c_{s}\thickapprox0$,
DE fluctuations are effectively pressureless and behave as matter
fluctuations. In this case, one can modify the SC model to include
this extra clustering component \cite{Abramo2007,Creminelli2010,Chang:2017vhs}.
However, for non-negligible $c_{s}$, DE fluctuations are affected
by their pressure gradients, and do not follow the matter evolution.
Therefore, the SC model has to be modified to account for this effect.
The first effort in this direction was made in \cite{Basse2011},
which included DE linear perturbations in the evolution of the SC
model.

Only in the last couple years, studies based on numerical N-body simulations
codes began to include DE fluctuations. In \cite{Dakin2019}, DE linear
perturbations are included as a source of the gravitational potential.
It was found that, even without nonlinear DE fluctuations, the matter
power spectrum can change at the percent level. In \cite{Hassani2019},
DE nonlinear fluctuations where treated, showing that they can indeed
become nonlinear and change the formation of halos for $c_{s}^{2}=10^{-7}$.
However, so far, this kind of studies did not yet consider the impact
of $c_{s}$ on halo mass functions. As we will show, in terms of virialization
threshold ($\delta_{{\rm v}}$) and density profiles, DE fluctuations
become effectively pressureless for $c_{s}^{2}<10^{-5}$ on small
nonlinear scales, but higher values also produce relevant changes
with respect to the nearly homogeneous case with $c_{s}^{2}=1$. 

In order to show these effects, we develop a method to solve the nonlinear
evolution of perfect fluids with spherical symmetry, particularly
in the case of pressureless matter and DE with arbitrary $w$ and
$c_{s}$. We present the equations in the Pseudo-Newtonian framework
and solve them numerically using the pseudo-spectral method with collocation
points. We show the impact of $c_{s}$ on the linear and nonlinear
evolution of matter fluctuations, matter peculiar velocities and gravitational
potential. We also compute the threshold of virialization for various
values of $c_{s}$, which can be used to estimate the impact of DE
fluctuations on the abundance of halos. 

Although this approach is not as realistic as a N-body simulation,
it is much more economical in computational power, allowing us to
understand the dependence of DE fluctuation on $c_{s}$, and not only
for very low values. A typical code run takes about 10 minutes in
an Intel i7 core, with very small memory use. The method can be easily
generalized for other models, like coupled DE-DM, warm DM, modified
gravity and Ultra Light DM. Therefore, we can easily explore model
parameters and predict which scenario is potentially observationally
distinguishable in view of current or future observations. This kind
of study can also be a guide to more realistic simulations. 

This paper is organized as follows. In section 2, we show the system
of equations that describe the fluctuations in the two fluids and
their initial conditions. In section 3, we present the numerical method
used to solve the resulting equations. In section 4, we show the impact
of $c_{s}$ on DM and DE profiles and discuss a pathology associated
with the nonlinear evolution of phantom models ($w<-1$). We calculate
the virialization threshold in section 5 and conclude in section 6.

\section{Equations of motion}

We will use the Pseudo-Newtonian Cosmology \cite{Lima:1996at} in
order to describe the evolution of pressureless matter and DE fluctuations
in the nonlinear regime, written in physical coordinates. For each fluid,  we have
the following equations: 
\begin{equation}
\frac{\partial\rho}{\partial t}+\vec{\nabla}\cdot\left(\vec{u}\rho\right)+p\vec{\nabla}\cdot\vec{u}=0\, ,
\end{equation}
\begin{equation}
\frac{\partial\vec{u}}{\partial t}+\left(\vec{u}\cdot\vec{\nabla}\right)\vec{u}=-\vec{\nabla}\Phi-
\frac{\vec{\nabla}p}{\rho+p}\, ,
\end{equation}
and for the gravitational potential we have
\begin{equation}
\nabla^{2}\Phi=4\pi G\sum_{j}\left(\rho_{j}+3p_{j}\right)\,.
\end{equation}
As usual, we split background and fluctuations quantities. Assuming
spherical symmetry, all quantities depend on time and the radial coordinate,
$r$, which we define as comoving with the background expansion: 
\begin{equation}
\rho=\bar{\rho}\left(t\right)+\delta\rho\left(t,r\right)\,,
\end{equation}
\begin{equation}
p=\bar{p}\left(t\right)+\delta p\left(t,r\right)\,,
\end{equation}
\begin{equation}
\vec{u}=\vec{u}_{0}+v\left(t,r\right)\hat{r}\,,
\end{equation}
\begin{equation}
\Phi=\Phi_{0}+\phi\left(t,r\right)\,.
\end{equation}
We also assume a time-dependent equation of state for the background
pressure, 
\begin{equation}
\bar{p}_{de}=w\left(t\right)\bar{\rho}_{de}\,,
\end{equation}
and a time-dependent sound speed, which relates the pressure and to
density fluctuations
\begin{equation}
\delta p_{de}=c_{s}^{2}\left(t\right)\delta\rho_{de}\,.
\end{equation}
Under these assumptions, the equations for the nonlinear evolution
of pressureless matter and DE are given by:
\begin{equation}
\dot{\delta}_{m}+\left(1+\delta_{m}\right)\frac{\partial_{r}\left(r^{2}v_{m}\right)}{ar^{2}}+\frac{v_{m}\partial_{r}\delta_{m}}{a}=0\label{eq:mat_rho}\,,
\end{equation}

\begin{equation}
\dot{v}_{m}+Hv_{m}+\frac{v_{m}\partial_{r}v_{m}}{a}=-\frac{\partial_{r}\phi}{a}\label{eq:mat_vel}\,,
\end{equation}
\begin{equation}
\dot{\delta}_{de}+3H\left(c_{s}^{2}-w\right)\delta_{de}+\left[1+w+\left(1+c_{s}^{2}\right)\delta_{de}\right]\frac{\partial_{r}\left(r^{2}v_{de}\right)}{ar^{2}}+\frac{v_{de}\partial_{r}\delta_{de}}{a}=0\label{eq:de_rho}\,,
\end{equation}

\begin{equation}
\dot{v}_{de}+Hv_{de}+\frac{v_{de}\partial_{r}v_{de}}{a}=-\frac{\partial_{r}\phi}{a}-\frac{c_{s}^{2}\partial_{r}\delta_{de}}{a\left[1+w+\left(1+c_{s}^{2}\right)\delta_{de}\right]}\label{eq:de_vel}\,,
\end{equation}
\begin{equation}
\left(\partial_{r}^{2}+\frac{2}{r}\partial_{r}\right)\phi=\frac{3a^{2}H^{2}}{2}\left[\Omega_{m}\delta_{m}+\Omega_{de}\left(1+3c_{s}^{2}\right)\delta_{de}\right]\label{eq:poiss_fluc}\,.
\end{equation}

For scales well above the sound horizon of DE, the term $c_{s}^{2}\partial_{r}\delta_{de}$
can be neglected, and equations (\ref{eq:de_vel}) and (\ref{eq:mat_vel})
are identical, showing that both fluids flow in the same way. Under
this assumption, clustering DE models have been studied in various
scenarios \cite{Nunes:2004wn,Manera:2005ct,Abramo2007,Pace2010,Creminelli2010,Batista:2013oca,Pace2014a,Velten2014a,Batista2017}.
The same problem was also studied for non-negligible sound speed,
but assuming that dark energy perturbations are linear \cite{Basse2011}.
The main achievement of our work is the development of a numerical
code capable of consistently solving the evolution of this type of
model for arbitrary values of sound speed.

In the following, we assume a background evolution with flat spatial
section, pressureless matter (baryons plus dark matter) and DE with
CPL equation of state \cite{Chevallier:2000qy,Linder2007} $w=w_{a}+\left(1-a\right)w_{a}$.
Thus, the Hubble function is given by:
\begin{equation}
H^{2}=H_{0}^{2}\left(\Omega_{m}^{0}a^{-3}+\left(1-\Omega_{m}^{0}\right)f\left(a\right)\right),
\end{equation}
where $f\left(a\right)=a^{-3\left(1+w_{0}+w_{a}\right)}\exp\left[3w_{a}\left(a-1\right)\right]$.
In all examples shown, we assume $\Omega_{m}^{0}=0.3$. For simplicity,
we also assume a constant $c_{s}$.

\subsection*{Initial conditions}

We set the initial conditions in the matter-dominated era ($z_{i}=99$),
making use of well-known analytical solutions in the linear regime,
for instance, see \cite{Abramo2009,Sapone2009}. We assume an initial
Gaussian profile for matter, which initially follows the EdS solution
\begin{equation}
\delta_{m}\left(a,r\right)=A\left(\frac{a}{a_{i}}\right)\exp\left(-\frac{r^{2}}{2\sigma^{2}}\right)\label{eq:mat_lin_EdS}\,.
\end{equation}
Then we determine the initial velocity profile for matter with the
linearized version of Eq. (\ref{eq:mat_rho}):
\begin{equation}
v_{m}\left(a_{i},r\right)=-\frac{a_{i}H_{i}}{r^{2}}\int_{0}^{r}dr'\left(r'\right)^{2}\delta_{m}\left(a_{i},r'\right)\,.\label{eq:vm_lin_EdS}
\end{equation}
The initial potential profile is determined assuming that, initially
$\Omega_{m}\delta_{m}\gg\Omega_{de}\delta_{de}$:
\begin{equation}
\left(\partial_{r}^{2}+\frac{2}{r}\partial_{r}\right)\phi=\frac{3H_{i}^{2}}{2}\Omega_{m}\left(a_{i}\right)\delta_{m}\left(a_{i},r\right)\,,\label{eq:poiss_IC}
\end{equation}
which has the analytical solution
\begin{equation}
\phi=-\frac{3H_{i}^{2}}{2}\Omega_{m}\left(a_{i}\right)A\sigma^{3}\sqrt{\frac{\pi}{2}}\frac{1}{r}\text{erf}\left(\frac{r}{\sqrt{2}\sigma}\right)\,.
\end{equation}

For DE, we implement two kinds of initial conditions, depending on
the value of $c_{s}$. If $c_{s}<c_{sd}$, where $c_{sd}$ is some
reference value below which DE fluctuations behave as dust, we have
\begin{equation}
\delta_{de}\left(a_{i},r\right)=\frac{1+w}{1-3w}\delta_{m}\left(a_{i},r\right)\label{eq:de_lin_IC_cs_Null}
\end{equation}
and 
\begin{equation}
v_{de}\left(a_{i},r\right)=v_{m}\left(a_{i},r\right)\,.
\end{equation}
For $c_{s}>c_{sd}$, DE perturbations are much smaller then matter
perturbations and the initial conditions are given by

\begin{equation}
\delta_{de}=-\frac{1+w}{c_{s}^{2}}\phi\label{eq:de_lin_IC}
\end{equation}
and 
\begin{equation}
v_{de}\left(a_{i},r\right)=-\frac{3a_{i}H_{i}\left(c_{s}^{2}-w\right)}{\left(1+w\right)r^{2}}\int_{0}^{r}dr'\left(r'\right)^{2}\delta_{de}\left(a_{i},r'\right)\,.\label{eq:de_vm_lin_IC}
\end{equation}
Off course, in the case $c_{s}\sim c_{d}$ these initial conditions
might not be quite satisfactory, but this is not an issue for the
late-time evolution, when transient behavior due to imprecise initial
conditions are usually negligible. As will be shown, for the small
nonlinear scales, DE models with $c_{s}^{2}<10^{-3}$ begin to deviate
significantly from the homogeneous case ($c_{s}^{2}=1$), thus we
assume $c_{sd}^{2}=10^{-3}$. 

However, in general, we observe that for $c_{s}^{2}>10^{-3}$ the
numerical evolution is more stable if we start with $\delta_{de}=0$.
Again, this choice has little effect on the late time evolution of
$\delta_{de}$. Moreover, in this case, we will show that DE fluctuations
are much smaller than matter fluctuations. Therefore, the reduced
accuracy in these nearly homogeneous DE scenarios has a negligible
effect on the late-time values of $\delta_{m}$ and $\phi$. The origin
of this issue is likely related to the boundary conditions for $\delta_{de}$
that will be discussed in the next section. They are chosen to better
represent models with low $c_{s}$, in which case the impact of DE
fluctuation is much more relevant. 

\section{Numerical method}

We numerically integrate the equations of motion (\ref{eq:mat_rho})-(\ref{eq:de_vel})
using the Galerkin-Collocation method \cite{Alcoforado2021}, which
is one variant of spectral methods. The starting point is the establishment
of approximate expressions for the relevant dynamical quantities $\delta_{m}(t,r)$,
$v_{m}(t,r)$, $\delta_{de}(t,r)$, $v_{de}(t,r)$ and $\phi(t,r)$:

\begin{eqnarray}
\delta_{m}(t,r) & = & \sum_{k=0}^{N}\,a_{k}(t)\psi_{k}(r),\;\;v_{m}(z,r)=\sum_{k=0}^{N-1}\,v_{m\,k}(z)\chi_{k}(r),\label{eq:mat_expan}\\
\delta_{de}(t,r) & = & \sum_{k=0}^{N}\,b_{k}(t)\psi_{k}(r),\;\;v_{de}(z,r)=\sum_{k=0}^{N-1}\,v_{de\,k}(z)\chi_{k}(r),\label{eq:de_expan}\\
\phi(t,r) & = & \sum_{k=0}^{N}\,\phi_{k}(t)\xi_{k}(r),\label{eq:phi_expan}
\end{eqnarray}

\noindent where $a_{k}(t),v_{m\,k}(t),b_{k}(t),v_{de\,k}(t)$ and
$\phi_{k}(t)$ are the unknown modes that constitute the spectral
representation of the respective dynamical quantities of interest;
$N$ is the truncation order that limits the number of terms in the
series expansion. The functions $\psi_{k}(r),\chi_{k}(r),\xi_{k}(r)$
are defined in the whole spatial domain, $0\leq r<\infty$, and expressed
as suitable combinations of the rational Chebyshev polynomials \cite{Boyd2001}
in order to satisfy the following boundary conditions,

\begin{eqnarray}
\delta_{m} & = & f(t)+\mathcal{O}(r),\;\;v_{m}=\mathcal{O}(r),\label{eq:dm_boundary}\\
\delta_{de} & = & g(t)+\mathcal{O}(r),\;\;v_{de}=\mathcal{O}(r),\label{eq:de_boundary}\\
\phi & = & \phi_{0}(t)+\mathcal{O}(r^{2}),
\end{eqnarray}

\noindent near $r=0$, and,

\begin{eqnarray}
\delta_{m},\delta_{e} & = & \mathcal{O}(r^{-1}),\;\;v_{m},v_{e}=\mathcal{O}(r^{-1}),\\
\phi & = & \mathcal{O}(r^{-1}),
\end{eqnarray}

\noindent valid at the spatial infinity, $r=\infty$.

The basis functions that satisfy the above boundary conditions are
defined as convenient linear combinations of the rational Chebyshev
polynomials, $TL_{k}(r)$, given by \cite{Boyd2001},

\begin{equation}
TL_{k}(r)\equiv T_{k}\left(x=\frac{r-L_{0}}{r+L_{0}}\right),
\end{equation}

\noindent where $T_{k}(x)$ represents the usual Chebyshev polynomials,
and $L_{0}$ is the map parameter. Accordingly, the basis functions
are defined as:

\begin{eqnarray}
\psi_{j}(r) & = & \frac{1}{2}(-1)^{j+1}(TL_{j+1}(r)-TL_{j}(r))\,,\\
\chi_{j}(r) & = & (-1)^{j+1}(\psi_{j+1}(r)-\psi_{j}(r))\,,\\
\xi_{j}(r) & = & -\frac{1+2j+2j^{2}}{3+2j+2(j+1)^{2}}\psi_{j+1}(r)+\psi_{j}(r)).
\end{eqnarray}

\noindent We remark that the requirement of the basis functions to
satisfy the boundary conditions is a typical feature of the Galerkin
method. On the other hand, we shall use a characteristic of the collocation
method, namely, the unknown modes are chosen such that the approximations
described by Eqs. (\ref{eq:mat_expan})-(\ref{eq:phi_expan}) coincide
with the corresponding exact functions at certain points, known as
the collocation or grid points. For instance, we can write the following
relation for the contrast density of matter:

\begin{equation}
\delta_{m}(t,r_{j})=\sum_{k=0}^{N}\,a_{k}(t)\psi_{k}(r_{j})\equiv\delta_{m[j]}^{\mathrm{(exact)}}(t)\,.
\end{equation}

\noindent The set of values of the density contrast of matter at the
collocation points $\delta_{m[j]}^{\mathrm{(exact)}}\left(t\right)$,
$j=0,1,..,N$, constitutes the physical representation of $\delta_{m}$
that is related to its correspondent spectral representation formed
by the coefficients $a_{k}$. The collocation points are given by,

\begin{eqnarray}
x_{j} & = & \cos\left(\frac{j\pi}{N}\right),\;\;j=0,1,..,N,\;\;\mathrm{and}\\
r_{j} & = & L_{0}\frac{1+x_{j}}{1-x_{j}}.
\end{eqnarray}

The interplay between both representations will be determinant for
an efficient implementation of the spectral algorithm to evolve the
field equations. In this way, by substituting the approximations
(\ref{eq:mat_expan})-(\ref{eq:phi_expan}) into the systems of equations
(\ref{eq:mat_rho})-(\ref{eq:poiss_fluc}), we generate the correspondent
residual equation. Following the collocation method, these equations
are forced to vanishes exactly at the collocation points. For the
sake of clearness, let us consider equation (\ref{eq:mat_vel}) in which
by imposing that the correspondent residual equation vanish at the
collocation points, it follows,

\begin{eqnarray}
\mathrm{Res}(t,r_{j})=\dot{v}_{m[j]} & + & Hv_{m[j]}+\frac{1}{a}v_{m[j]}\,\sum_{k=0}^{N}\,vm_{k}(z)\chi_{k}^{\prime}(r_{j})+\nonumber \\
 & + & \frac{1}{a}\,\sum_{k=0}^{N}\,\phi_{k}(z)\xi_{k}^{\prime}(r_{j})=0,
\end{eqnarray}

\noindent for all $j=0,1,..,N$. Notice that $\dot{v}_{m[j]}$ and
$\phi_{[j]}$ are the values of $\dot{v}_{m}$ and $\phi$ evaluated
at the collocation points. Repeating this procedure to the remaining
equations, we end up with a coupled system of ordinary differential
equations for the values of $\dot{\delta}_{m},\dot{v}_{m},\dot{\delta}_{e}$
and $\dot{v}_{e}$. We solve the system of ordinary differential equations
using the Gnu Scientific Library routines implemented in C language.

\section{Evolution of the profiles}

Let's study some qualitative aspects of the nonlinear evolution of
matter and DE fluctuations. Excluding models with phantom crossing,
we analyze the evolution of matter overdensities and underdensities
for non-phantom ($w>-1$) and phantom ($w<-1$) DE. Given the approximate
solutions (\ref{eq:de_lin_IC}) and (\ref{eq:de_lin_IC_cs_Null}),
we expect that matter fluctuations will be correlated with DE fluctuations
with non-phantom EoS, while for phantom EoS DE fluctuations should
be anti-correlated with those of matter. For all the examples shown,
the initial matter profile has $\sigma=30\text{Mpc/h}$.

\subsection{Linear evolution}

Let's first consider the linear evolution of matter and DE perturbations.
Although this task can be efficiently done in the Fourier space, it's
important to check how our implementation performs for our specific
profile. In Appendix A, we present a convergence and accuracy study
for the evolution of profiles in EdS model, showing that we can achieve
errors smaller than $10^{-6}\%$ for $\delta_{m}$ in the central
regions. 

\begin{figure}
\centering{}\includegraphics[scale=0.6]{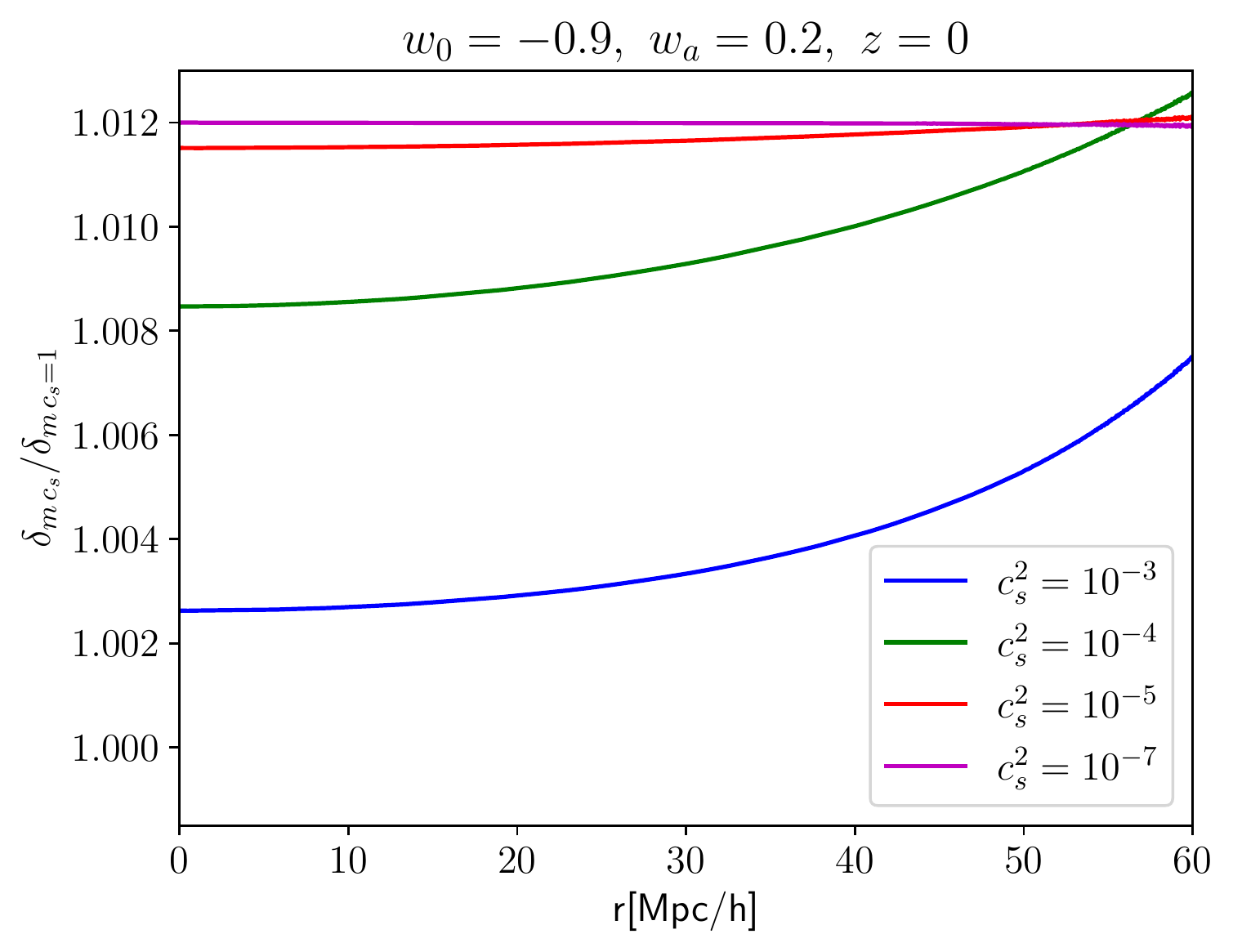}\caption{Profiles ratios $\delta_{m\,,c_{s}}/\delta_{m\,,c_{s}=1}$ at $z=0$
for selected values of $c_{s}$. The initial conditions for $\delta_{m}$
are the same in all cases. \label{fig:dm-linear}\protect \\
}
\end{figure}

In figure \ref{fig:dm-linear} we plot the ratio of the linearly evolved
matter profiles to the profile with $c_{s}=1$, $\delta_{m\,,c_{s}}/\delta_{m\,,c_{s}=1}$.
For $c_{s}=1$, DE perturbations are negligible on small scales, even
in the nonlinear regime (see figure \ref{fig:halos_dm_de}). Therefore,
the growth of $\delta_{m}$ is effectively scale-invariant. We verified
this behavior by observing that 
\begin{equation}
\delta_{m\,,c_{s}=1}\left(z=0,r\right)-A_{r}\delta_{m\,,c_{s}=1}\left(z=99,r\right)\sim10^{-5}\,,
\end{equation}
where the quantity
\begin{equation}
A_{r}=\frac{\delta_{m\,,c_{s}=1}\left(z=0,r=0\right)}{\delta_{m\,,c_{s}=1}\left(z=99,r=0\right)}\,
\end{equation}
rescales the profile with the matter growth computed at the center.
Therefore, radial deviations from the profile $\delta_{m\,,c_{s}=1}\left(r,z\right)$
indicate scale-dependent growth, which is expected for lower values
of $c_{s}$. As seen in figure \ref{fig:dm-linear}, this clearly
happens for $c_{s}^{2}=10^{-3}$ and $c_{s}^{2}=10^{-4}$. For $c_{s}^{2}\le10^{-5}$,
the growth is also nearly scale-independent because DE perturbations
tend to behave as dust. In all cases, a lower sound speed enhances
the matter growth compared to the $c_{s}=1$ case.

\subsection{Matter halos\label{subsec:Matter-halos}}

Now we analyze the impact of $c_{s}$ on matter profiles associated
with the formation of halos. Starting with the same initial conditions
for matter fluctuations at $z_{i}=99$, we show the profiles at very
low-$z$. The value of $\delta_{m}\left(r=0,z_{i}\right)$ is chosen
to produce a profile that roughly represents virialization overdensities
($\delta_{m}\sim200$) at $z=0.04$. 

\begin{figure}
\centering{}\includegraphics[scale=0.5]{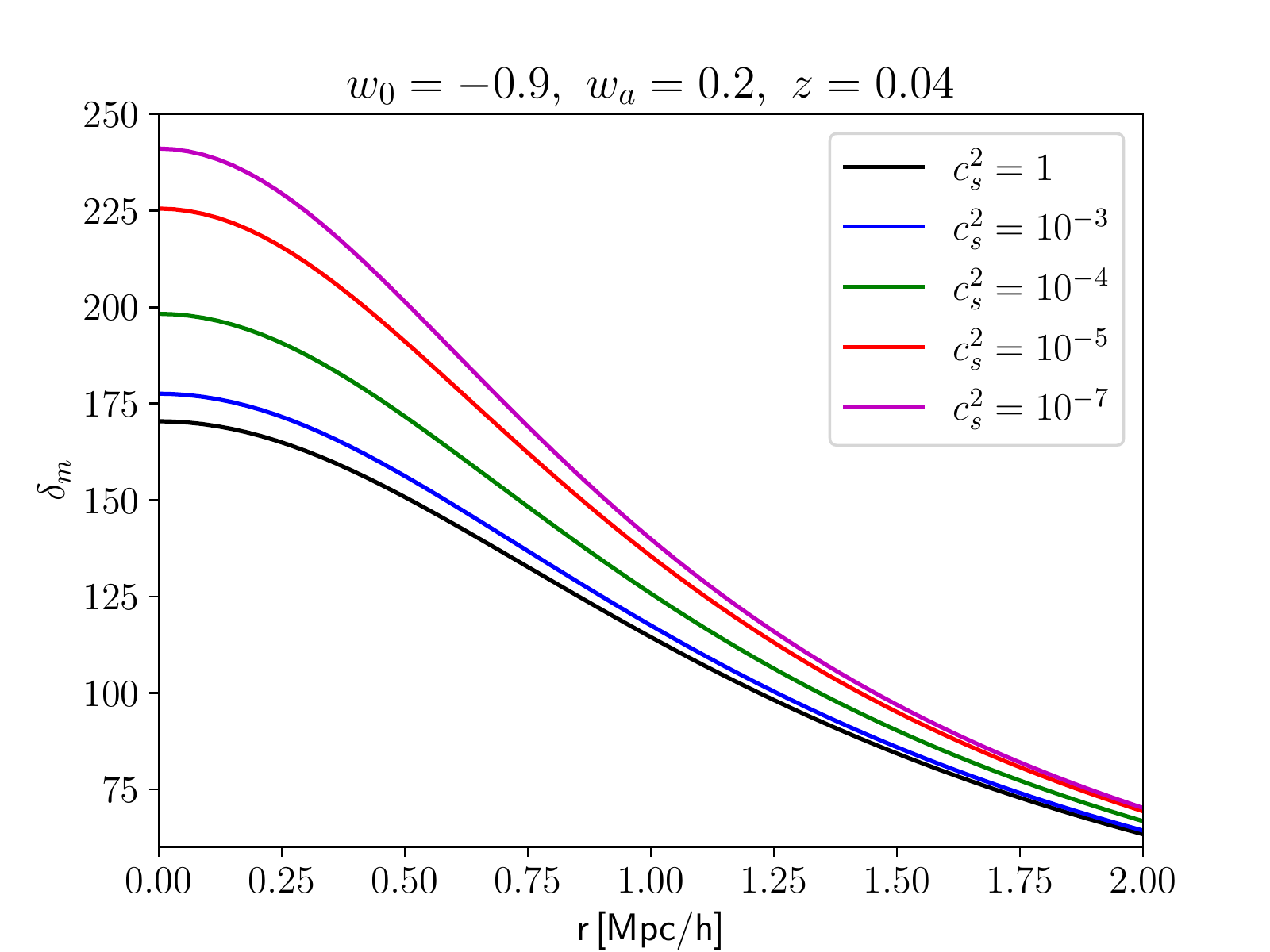}\includegraphics[scale=0.5]{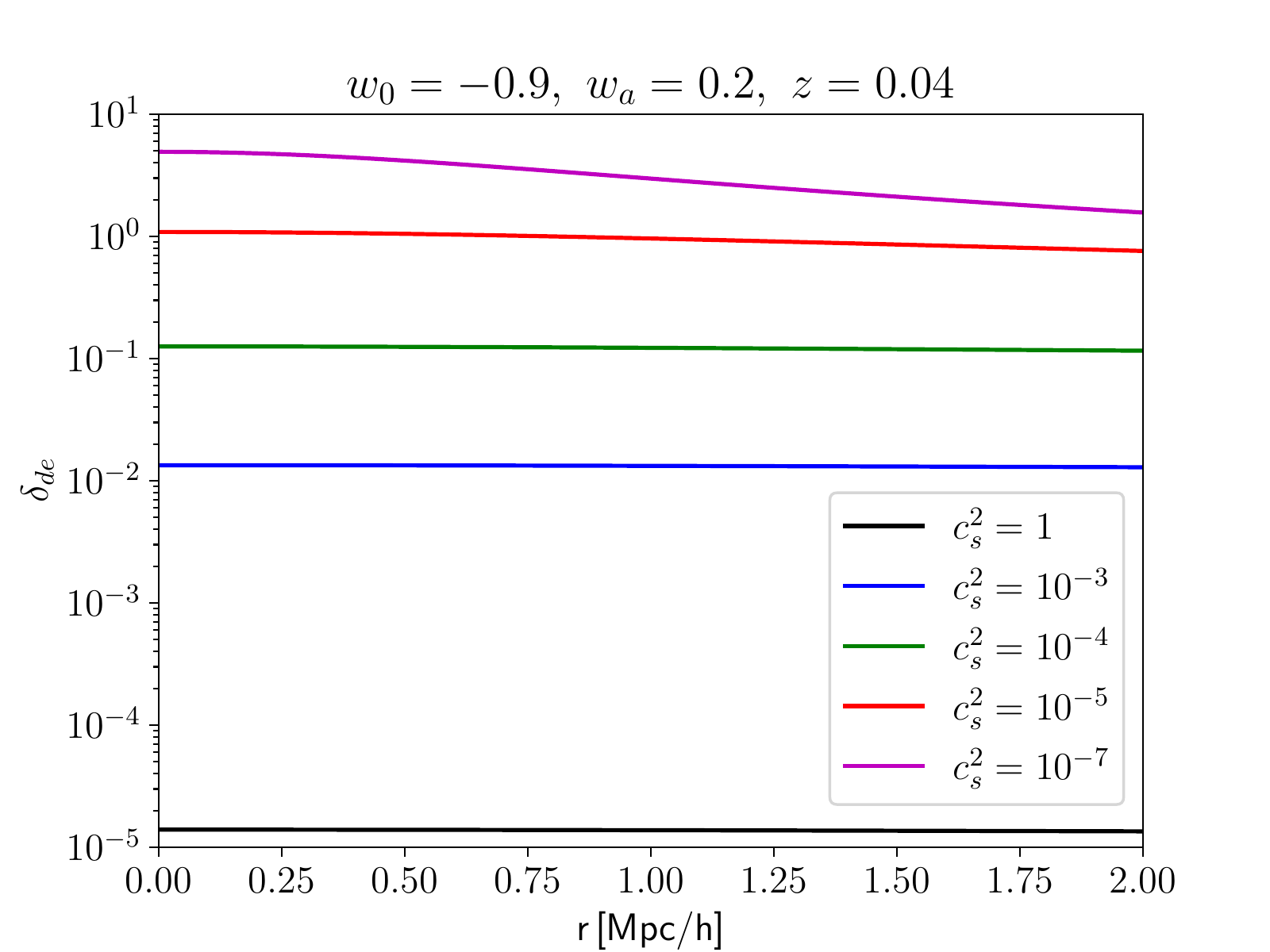}\caption{Left panel: impact of DE fluctuations on the nonlinear matter profile
at $z=0.04$ for selected values of $c_{s}$. Right panel: profiles
of $\delta_{de}$ for the corresponding cases show in the left panel.
The initial conditions for matter fluctuations are the same in all
cases. \label{fig:halos_dm_de}\protect \\
}
\end{figure}

In the left panel of figure \ref{fig:halos_dm_de} we can see that
lower values of $c_{s}$ enhances matter clustering. For $c_{s}^{2}>10^{-3}$,
this enhancement is small when compared to $c_{s}^{2}=1$. For $c_{s}^{2}\le10^{-7}$,
we verified that $\delta_{m}$ barely changes. The range of variation
of the central value of $\delta_{m}$ is substantial, showing that
even a small contribution of DE can produce a large modification in
the matter fluctuation in the nonlinear regime. 

In left panel of figure \ref{fig:halos_dm_de}, we show the corresponding
profiles of $\delta_{de}$ at $z=0.04$. For $c_{s}^{2}=1$, DE fluctuations
stay in the linear regime and are 7 orders of magnitude smaller than
matter fluctuations, which can be assumed as homogeneous DE on small
scales. In the case of $c_{s}^{2}=10^{-7}$, $\delta_{de}$ can reach
few percent of the corresponding matter fluctuations. For the scales
under consideration, we see that DE fluctuations become nonlinear
for $c_{s}^{2}<10^{-4}$. 

\begin{figure}
\centering{}\includegraphics[scale=0.6]{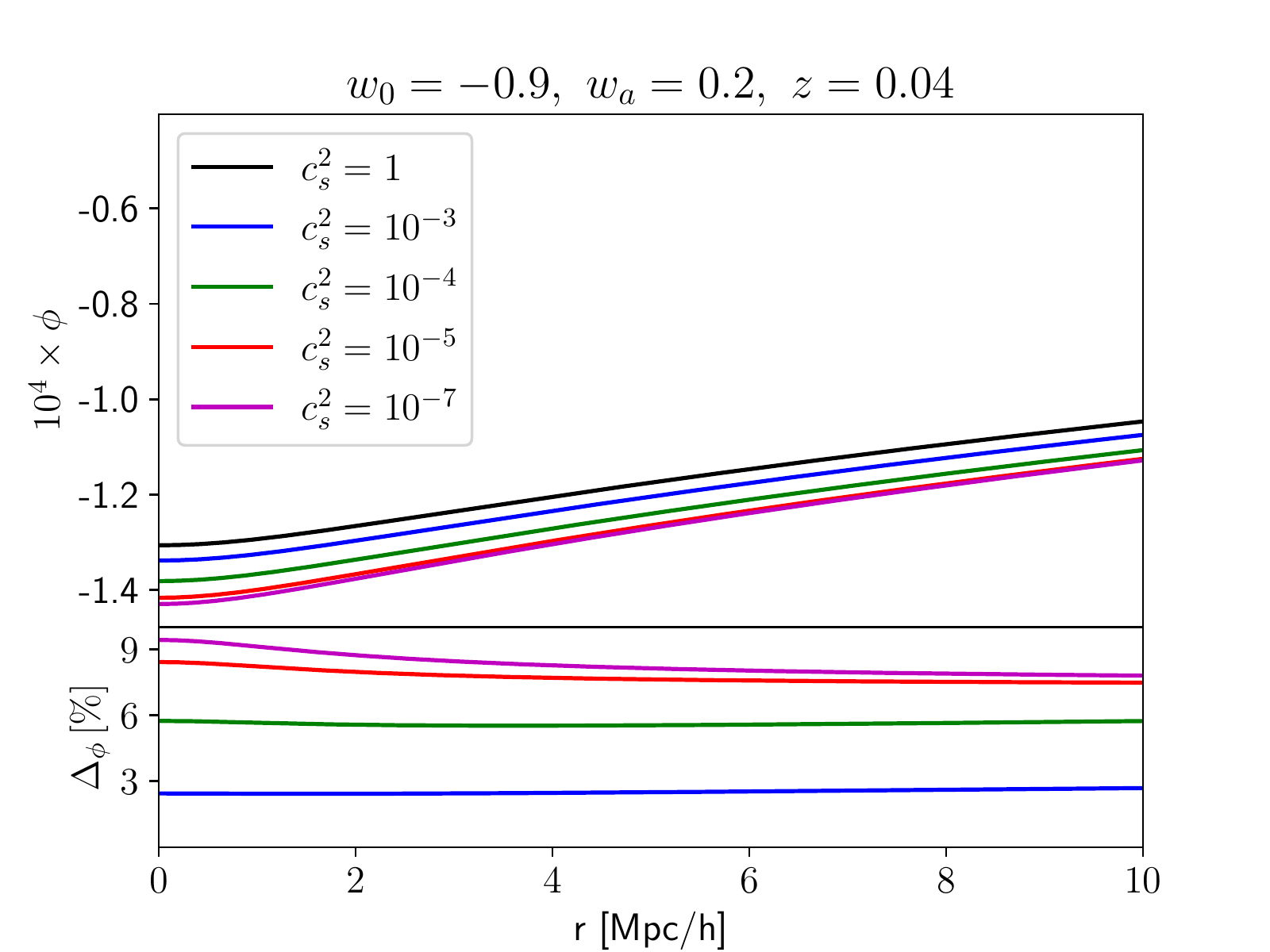}\caption{Top panel: Profiles of the potential $\phi$ at $z=0.04$ for selected
values of $c_{s}$. Lower panel: percent differences of the potential
for various $c_{s}$ with respect to the $c_{s}=1$ case, given by
$\Delta_{\phi}=100\times(\phi_{c_{s}}/\phi_{c_{s}=1}-1)$. The initial
conditions for $\delta_{m}$ are the same in all cases. \label{fig:phi_cs}\protect \\
}
\end{figure}

In figure \ref{fig:phi_cs}, we also show the impact of $c_{s}$ on
the gravitational potential. In the top panel we show $10^{4}\times\phi$
and in the lower panel the percent differences with respect to the
$c_{s}^{2}=1$ case, given by $\Delta_{\phi}=100\times\left(\frac{\phi_{c_{s}}}{\phi_{c_{s}=1}}-1\right)$.
As we can see, in the central region, the potential can change about
$10\%$ with respect to the homogeneous case ($c_{s}=1$). The impact
of $c_{s}$ is also present far away from the center, but is slightly
reduced. 

\begin{figure}
\centering{}\includegraphics[scale=0.6]{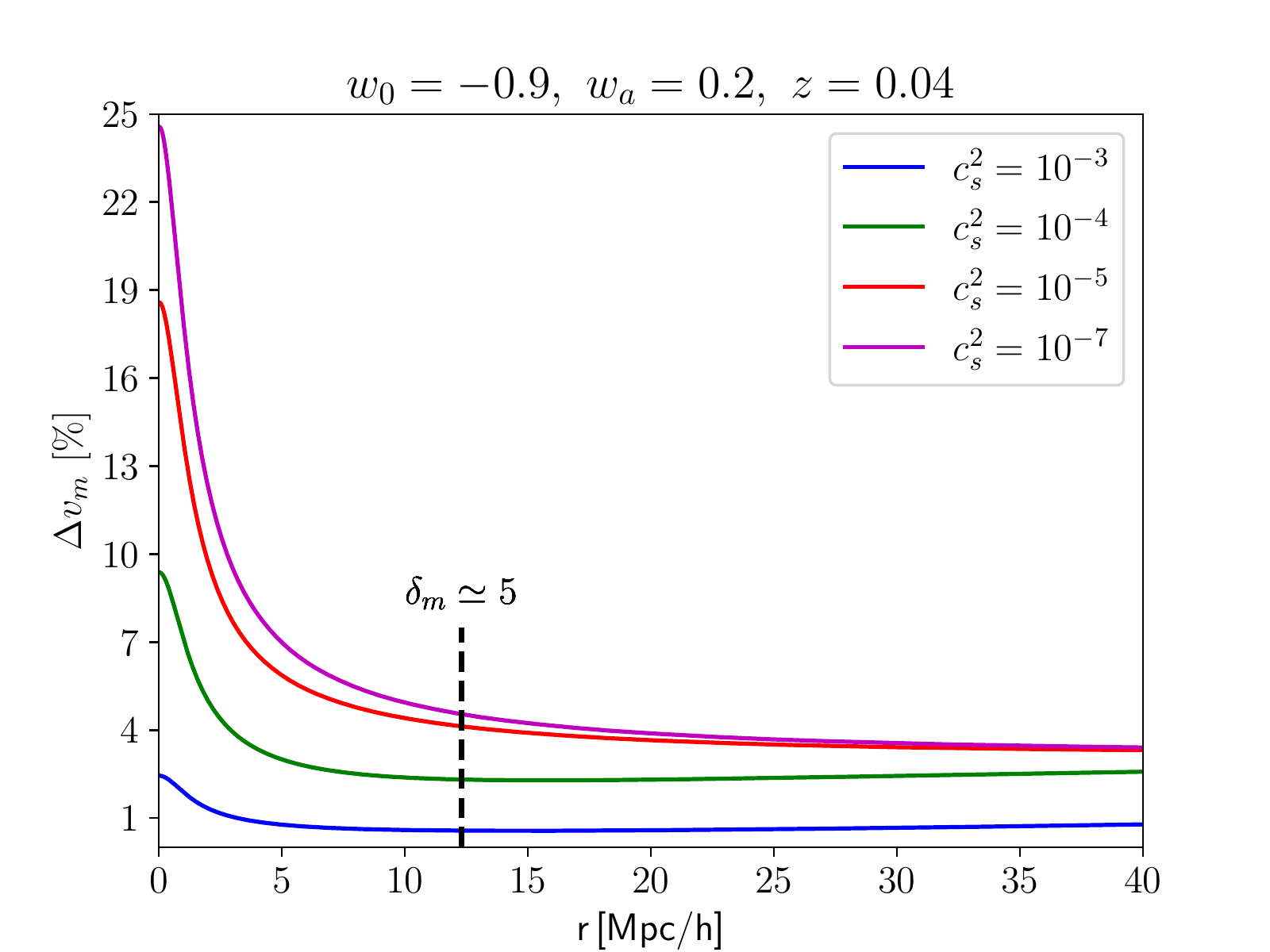}\caption{Change of the peculiar matter velocity, $v_{m}$, at $z=0.04$ for
selected values of $c_{s}$ with respect to the $c_{s}=1$ case, given
by $\Delta v_{m}=100\times\left(\frac{v_{mc_{s}}}{v_{mc_{s}=1}}-1\right)$.
The initial conditions for $\delta_{m}$ are the same as those used
in figure \ref{fig:halos_dm_de}. The vertical dashed-black line
shows the region where $\delta_{m}\simeq5$, which roughly indicates
the transition between collapsing and expanding regions. \label{fig:vm-cs}\protect \\
}
\end{figure}

We also check the impact of DE fluctuation of the peculiar matter
velocity, which is related to the redshift space-distortion effect.
In figure \ref{fig:vm-cs} we plot the percent difference of $v_{m}$
with respect to the case with homogeneous DE ($c_{s}=1$), given by
$\Delta v_{m}=100\times\left(\frac{v_{mc_{s}}}{v_{mc_{s}=1}}-1\right).$
The vertical dashed line indicates the radius such that $\delta_{m}\simeq5$.
This value roughly indicates the transition between collapsing nonlinear
and still expanding linear regions. In the nonlinear regions, DE fluctuations
can change $v_{m}$ substantially, in $10-20\%$ range. In the linear
regions, the variation with respect to the homegeneous case is only
about $3\%$ for the two lowest $c_{s}$ values. 

\subsubsection*{Phantom negative energy density}

In the previous examples, we used $w>-1$ for all the evolution. Now
let's analyze the case of a phantom equation of state. As already
noticed in the literature \cite{Abramo2007,Sapone2009,Creminelli2010},
in the limit $c_{s}\rightarrow0$, positive matter fluctuations will
induce negative phantom DE fluctuations, because $\delta_{de}\propto\left(1+w\right)\delta_{m}\,.$
Therefore, it is possible that matter halos can generate $\delta_{de}<-1$,
which is associated with the pathological situation of negative total
energy in the DE component $\rho_{de}=\bar{\rho}_{de}\left(1+\delta_{de}\right)$. 

\begin{figure}
\centering{}\includegraphics[scale=0.5]{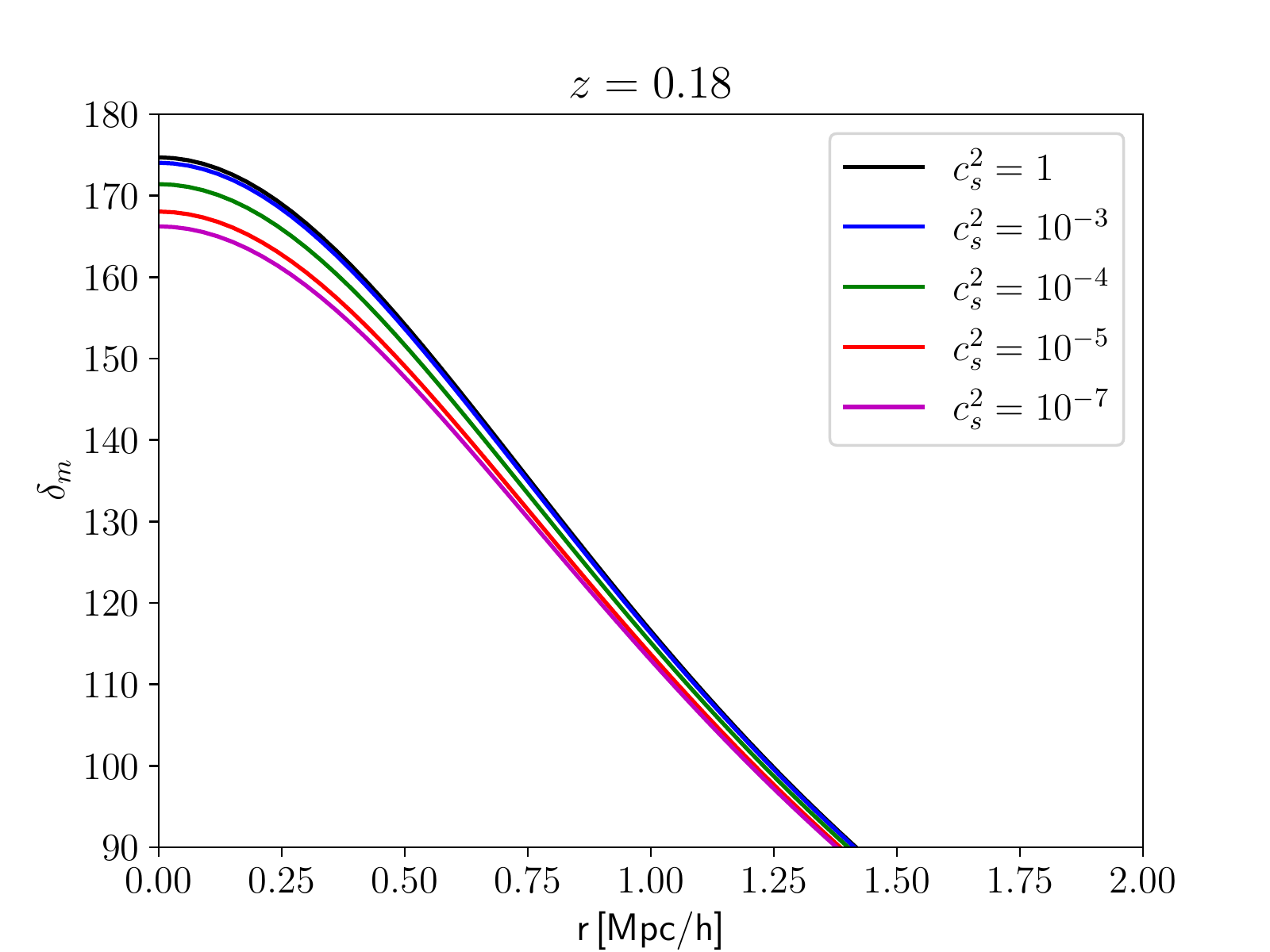}\includegraphics[scale=0.5]{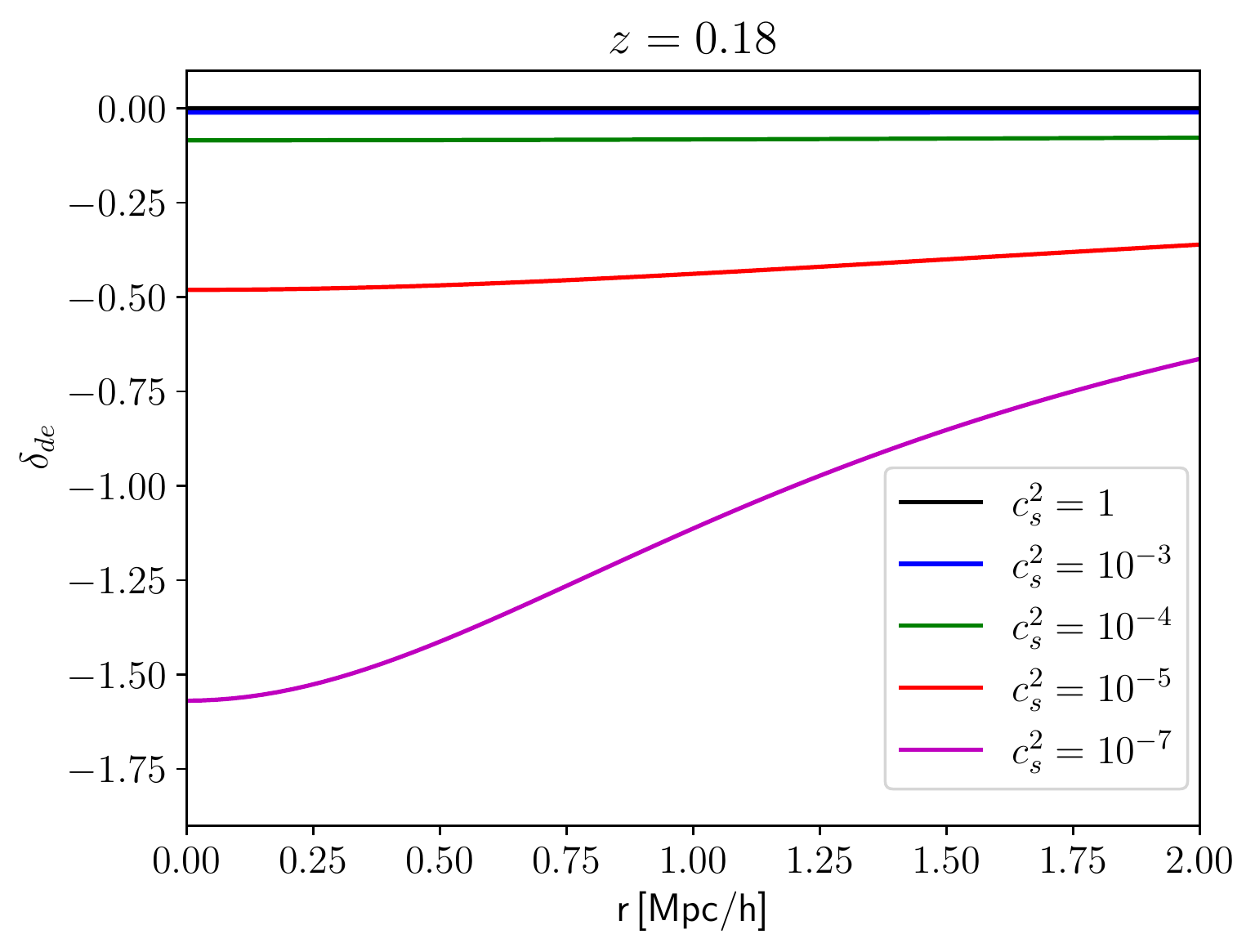}\caption{Left panel: $\delta_{m}$ profiles for $w=-1.1$ and selected values
of $c_{s}$ at $z=0.18$. Right panel: corresponding $\delta_{de}$
profiles. The initial conditions for $\delta_{m}$ are the same as
those used in figure \ref{fig:halos_dm_de}. \label{fig:phantom-profile}\protect \\
}
\end{figure}

In figure \ref{fig:phantom-profile}, for $w_{0}=-1.1$ and $w_{a}=0$,
we show that this situation is achieved by models with sufficiently
low sound speed. Note that, in these examples, $\delta_{m}$ roughly
presents virialization values at the central regions. This can be
understood as an averaged density contrast for the real halo profile.
The changes in the potential with respect to the homogeneous case
are smaller and opposite to the non-phantom case, reaching $-2.5\%$
for $c_{s}^{2}=10^{-7}$.

\begin{figure}
\centering{}\includegraphics[scale=0.6]{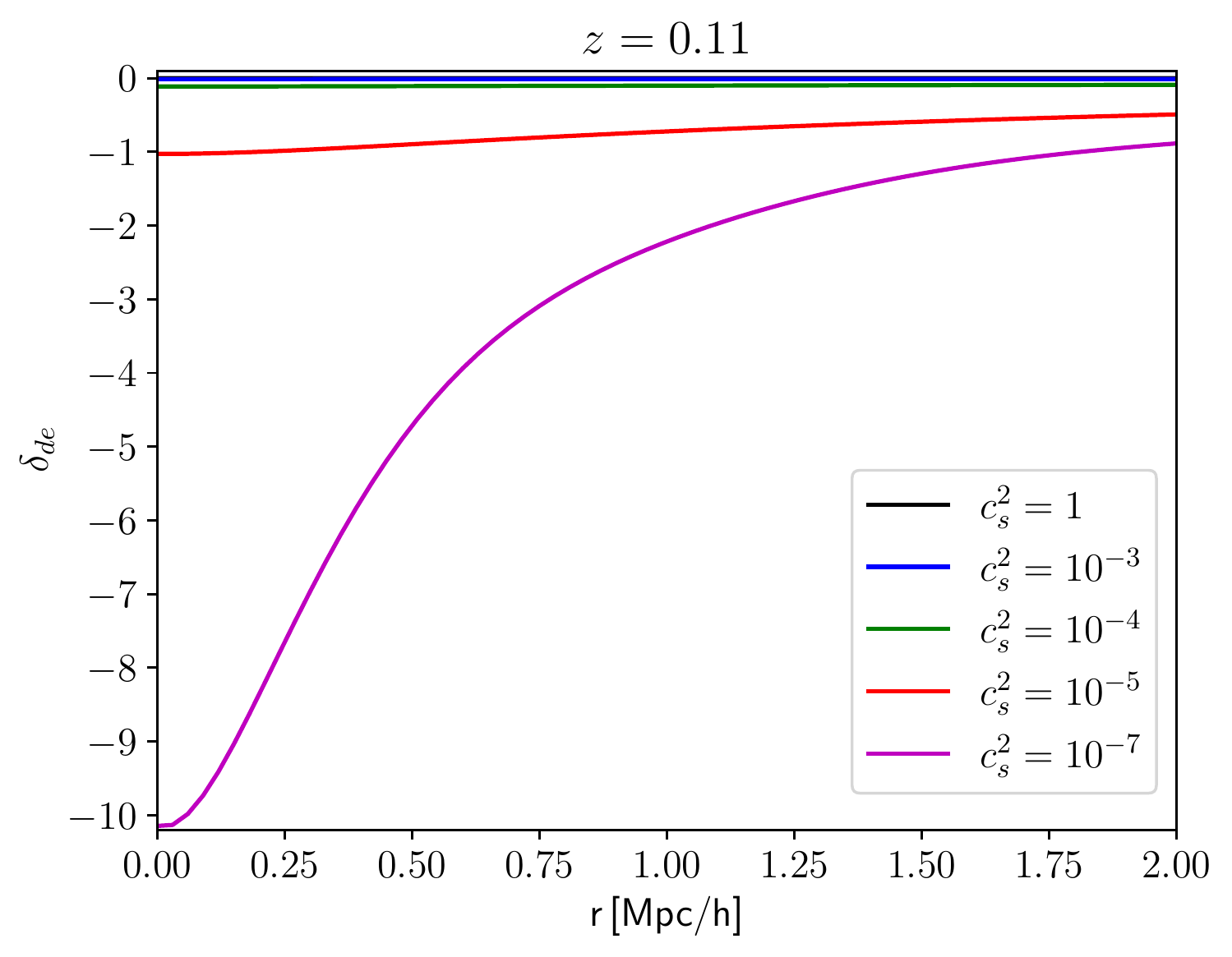}\caption{DE profiles for $w=-1.1$ at $z=0.11$ for selected values of $c_{s}$.
Here the matter profile is evolved up to $\delta_{m}\simeq1500$ at
the center. In this example, when $c_{s}^{2}<10^{-5}$ we have $\delta_{de}<-1$
around the center of the halo. \label{fig:de-phantom-profile-deep}\protect \\
}
\end{figure}

More realistic halo profiles can have $\delta_{m}\sim10^{3}$ at the
central region. Then, to avoid $\delta_{de}<-1$, larger $c_{s}$
is necessary. In figure \ref{fig:de-phantom-profile-deep} we show
the profiles for $\delta_{de}$ at $z=0.11$, but now with $\delta_{m}\simeq1500$
at the center. As can be seen, $\delta_{de}<-1$ is now achieved also
for $c_{s}^{2}=10^{-5}$. The DE contrast can be even more negative
for lower sound speed values. It is important to note that, having
in mind that $\delta_{de}\propto\left(1+w\right)\delta_{m}$, larger
values of $c_{s}$will be needed for more negative $w$ to avoid this
pathological behavior. 

The main driver of this pathological behavior is the term $1+w+\left(1+c_{s}^{2}\right)\delta_{de}$
in equation (\ref{eq:de_rho}), which couples the density contrast
to the gravitational potential. The corresponding term for matter
fluctuations is $1+\delta_{m}$, thus, when $\delta_{m}\rightarrow-1$,
the fluctuations decouple from $\phi$, the decrease of $\delta_{m}$
halts, and we always have $\delta_{m}>-1$. However, in general, this
coupling term for DE does not vanish when $\delta_{de}\rightarrow-1$,
and situations with $\delta_{de}<-1$ can be achieved for sufficiently
small $c_{s}$.

At face value, these phenomenological models present pathologies that
must be absent in any fundamental theory. Our examples demonstrate
that phantom models can not be described as perfect fluids with arbitrarily
low $c_{s}$, as in \cite{Creminelli2009}. In practice, many scalar
field models with $w<-1$ have no perfect fluid correspondence \cite{Pujolas2011,Sawicki2013a},
and dissipative effects may avoid this kind of problem in the nonlinear
regime.

\subsection{Matter voids}

\begin{figure}
\centering{}\includegraphics[scale=0.5]{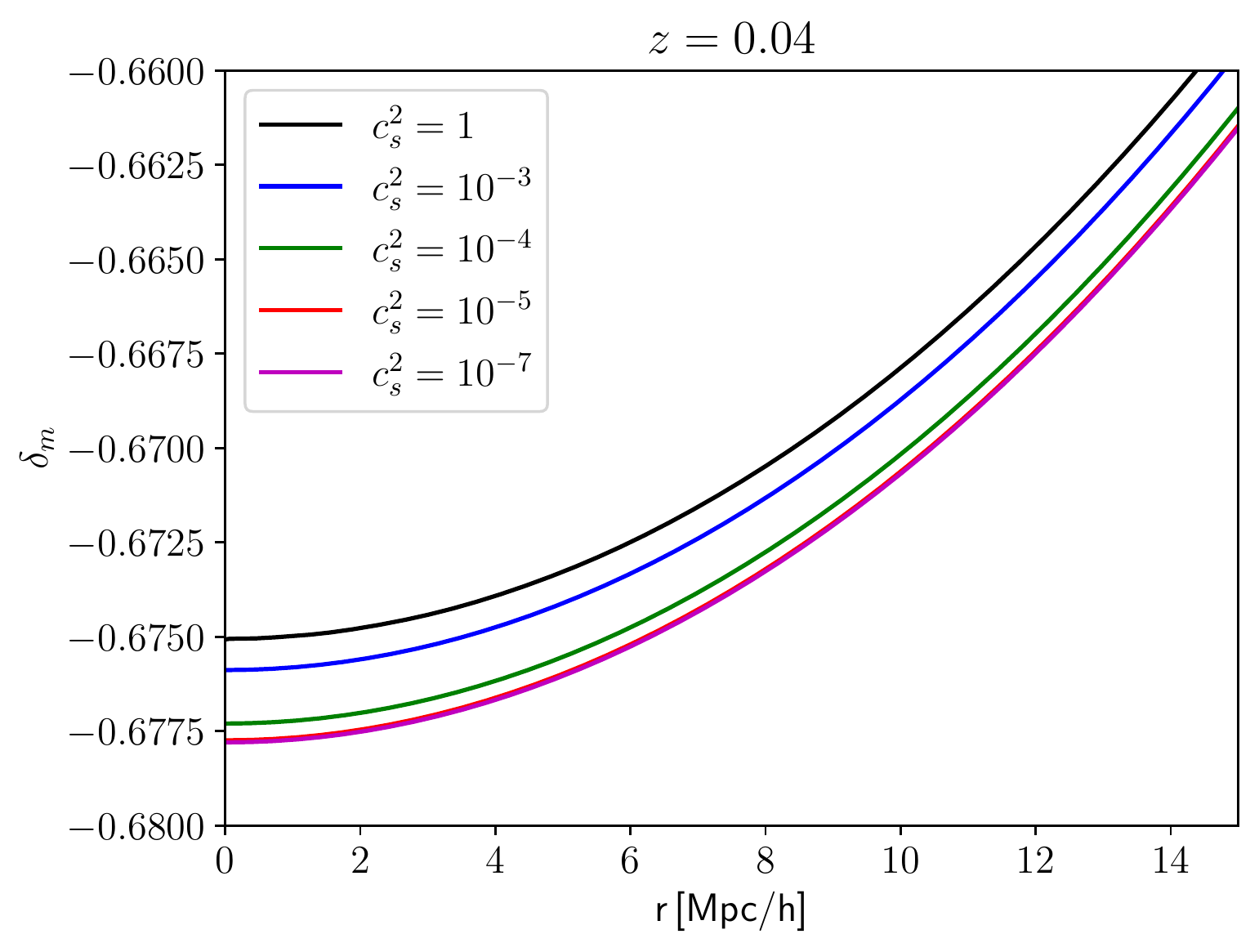}\includegraphics[scale=0.5]{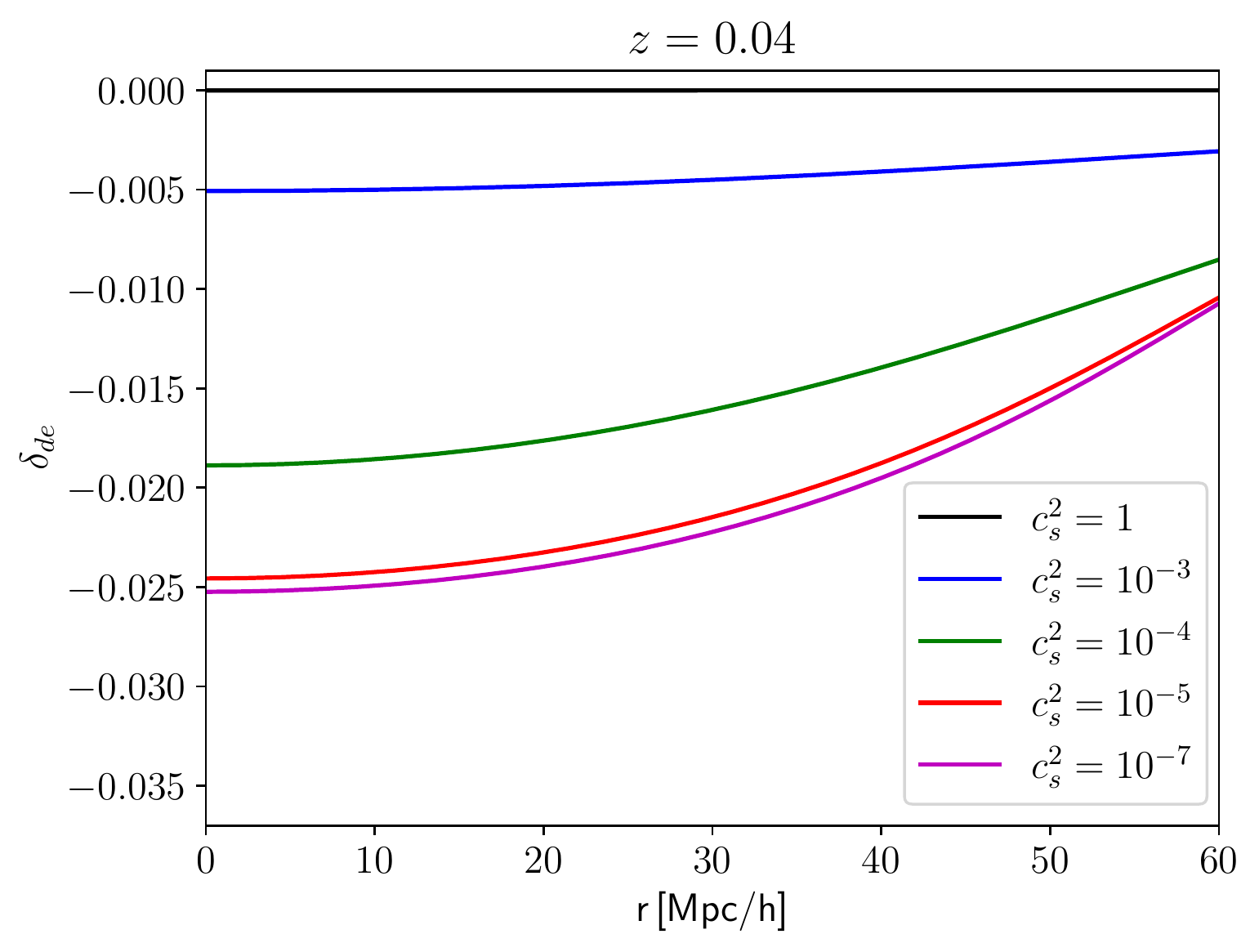}\caption{Left panel: impact of DE fluctuations on the nonlinear void matter
profile at $z=0.04$ for selected values of $c_{s}$. Right panel:
profiles of $\delta_{de}$ for the corresponding cases shown in the
left panel. The initial conditions for matter fluctuations are the
same in all cases. In the left panel, we focus on more central regions
so that the small differences (bellow than $1\%$) in the matter profiles
can be visible \label{fig:void-dm-de}\protect \\
}
\end{figure}

Let us estimate the impact of DE fluctuation on voids. Assuming the
same initial conditions as those used in figure \ref{fig:halos_dm_de},
but with negative values for $\delta_{m}$, we evolve the profiles
up to $z=0.04$. As can be seen in figure \ref{fig:void-dm-de},
the same kind of initial conditions that generate overdensities of
nearly virialized halos produce voids with $\delta_{m}\simeq-0.67$
at the central region. The impact of $c_{s}$ on $\delta_{m}$ is
much smaller for a void, below $1\%$. The variation of $\delta_{de}$
with $c_{s}$ is also smaller than in the case for halos.

We note that, in the left panel of figure \ref{fig:void-dm-de}, we
have $\delta_{de}>0$ for the $c_{s}=1$ case, following the approximate
solution $\delta_{de}\propto-\left(1+w\right)\phi/c_{s}^{2}$ for
models with relevant pressure support on small scales. For the cases
with $c_{s}^{2}<10^{-3}$, we have negative DE fluctuations, following
the dust-like approximate solution $\delta_{de}\propto\left(1+w\right)\delta_{m}$. 

\begin{figure}
\centering{}\includegraphics[scale=0.65]{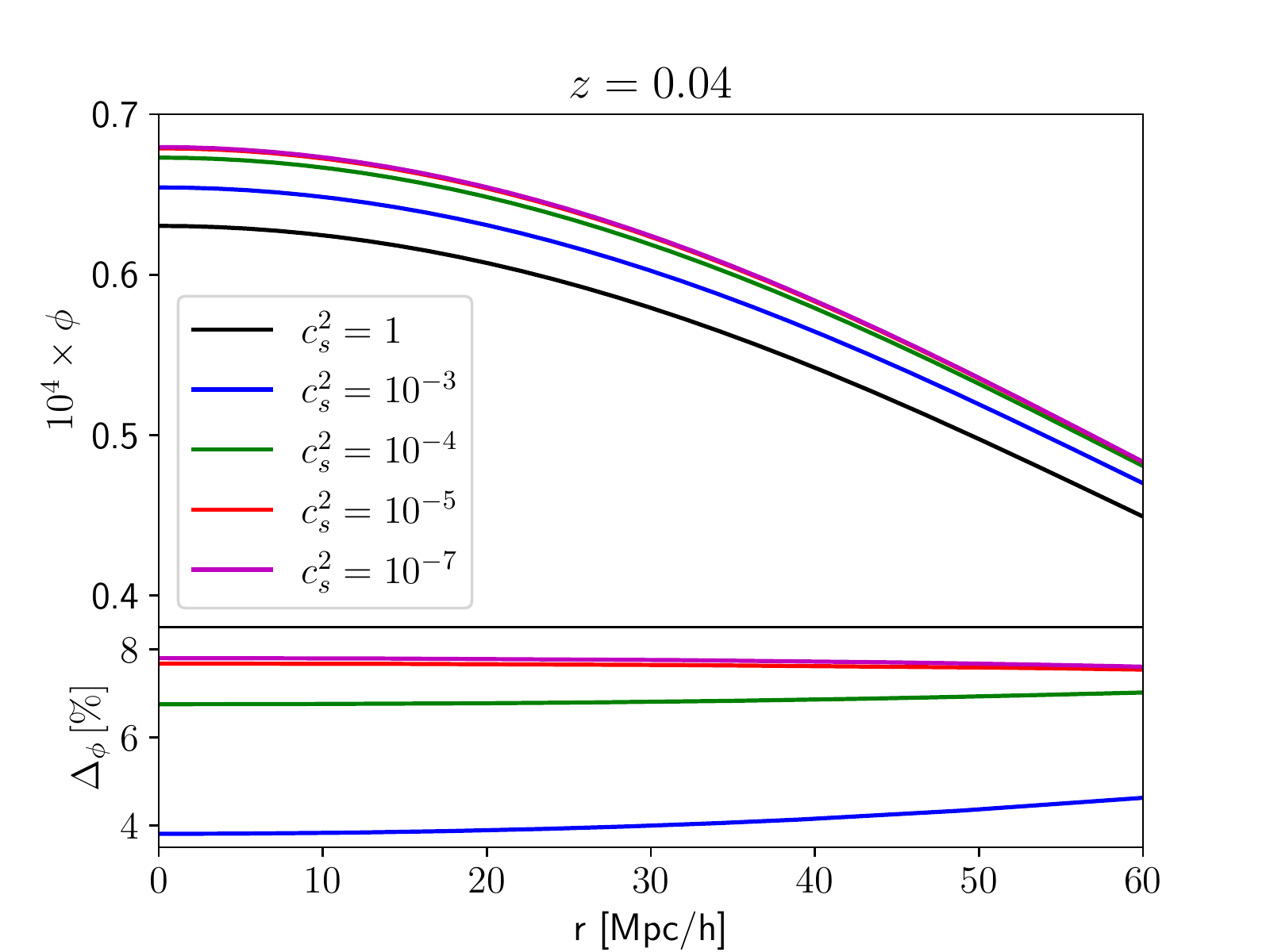}\caption{Top panel: Profiles of the potential $\phi$ at $z=0.04$ for selected
values of $c_{s}$. Lower panel: percent differences of the potential
for various $c_{s}$ with respect to the $c_{s}=1$ case, given by
$\Delta_{\phi}=100\times(\phi_{c_{s}}/\phi_{c_{s}=1}-1)$. The initial
conditions for $\delta_{m}$ are the same in all cases. \label{fig:phi-void-profile}\protect \\
}
\end{figure}

Although the impact of DE fluctuation in matter voids is smaller,
the change in the potential is similar to what we observed for halos.
In the lower panel of \ref{fig:phi-void-profile}, we see that $\phi$
can change about $8\%$ with respect to the homogeneous case. 

\subsection{Local DE EoS}

When DE fluctuations are non-negligible, it's local EoS, defined by
\begin{equation}
w_{c}=\frac{\bar{p}_{de}+\delta p_{de}}{\bar{\rho}_{de}+\delta\rho_{de}}=w+\left(c_{s}^{2}-w\right)\frac{\delta_{de}}{1+\delta_{de}}\,,\label{eq:w-loc}
\end{equation}
is expected to vary near a nonlinear structure \cite{Mota:2004pa,Abramo2007}.
With our method to solve for the profiles, we can now analyze how
$w_{c}$ changes in space. 

\begin{figure}
\centering{}\includegraphics[scale=0.65]{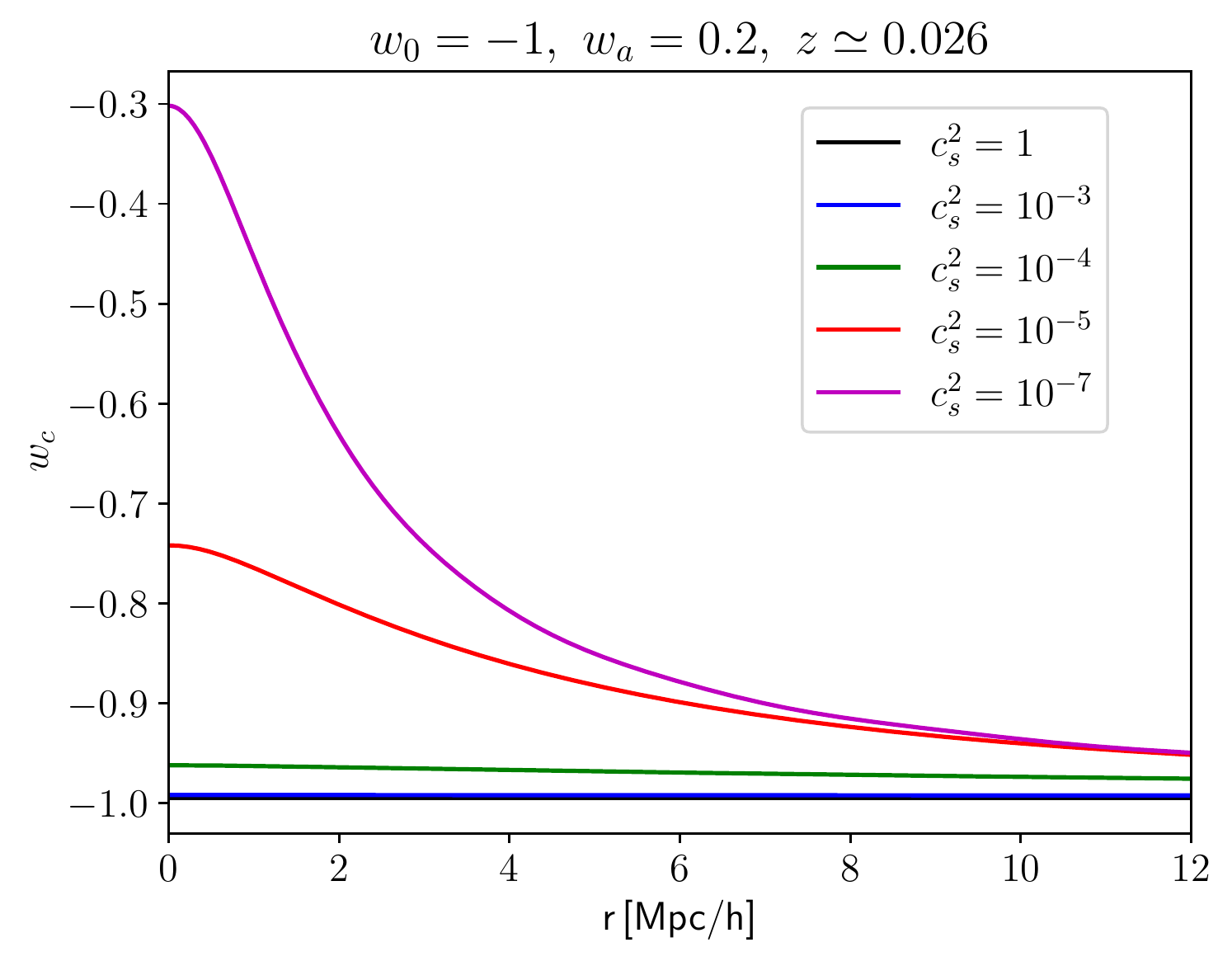}\caption{Profiles of the local DE EoS, $w_{c}$, at $z\simeq0.026$ for selected
values of $c_{s}$. The initial conditions for $\delta_{m}$ are the
same in all cases and roughly produce virialization overdensities
at the central regions at $z\simeq0.026$. \label{fig:w-local}\protect \\
}
\end{figure}

For this purpose, we choose $w_{0}=-1$ and $w_{a}=0.2$. This model
gives an EoS that is close to $-1$ at low$-z$, but is less negative
in the past, allowing DE fluctuation to grow and be present up to
now. In this case we have a background similar to $\Lambda$CDM at
low redshift, but with still relevant DE fluctuations. In figure \ref{fig:w-local},
we show the profile of $w_{c}$ at $z\simeq0.026$ using initial conditions
for $\delta_{m}$ such that its value at the center roughly represents
virialization values. As can be seen, for $c_{s}^{2}<10^{-5}$, the
change of the local equation of state with respect to $w$ can be
large near the center and still relevant in outer regions. In the
case of voids, we have $w_{c}\simeq w$ because the fluctuations of
DE are still linear. 

In the central regions, the local gravity of the halo dominates over
the background expansion. Thus the values of $w_{c}$ shall have a
negligible effect on light propagation and particle dynamics. But
in the outskirts of halos or in mildly nonlinear structures, it's
possible that departures of $w$ due to DE fluctuations can produce
a non-negligible effect. Such impact, however, depends crucially on
the actual matter distribution.

For phantom DE with low $c_{s}$ in the presence of a matter halo,
the local equation of state is ill-defined because $w_{c}$ diverges
when $\delta_{de}\rightarrow-1$. For healthy phantom models, it would
be possible to find a $c_{s}^{2}$ such that the change in $w_{c}$
is large. For instance, with $w_{0}=-1$, $w_{a}=-0.2$ and $c_{s}^{2}=10^{-5}$
we can find $\delta_{de}\left(r=0\right)\simeq-0.3$ at low-$z$,
which produces $w_{c}\left(r=0\right)\simeq-1.57$. However, this
kind of model is much more speculative because one would need to fine-tune
$c_{s}^{2}$ for each EoS under consideration to avoid $\delta_{de}<-1$. 

\section{Virialization threshold }

In the classical model for the spherical collapse in an EdS universe,
the evolution of $\delta_{m}$ can be solved analytically \cite{Gunn:1972sv,Padmanabhan}.
The top-hat nonlinear density diverges when the shell radius goes
to zero, which determines the redshift of collapse $z_{c}$, which,
in turn, is used to compute the critical density for collapse, $\delta_{c}$,
as the value of the linear evolved contrast at $z_{c}$. As well-known,
in EdS model, $\delta_{c}\simeq1.686$ is independent of redshift and
scale. In $\Lambda$CDM model, $\delta_{c}$ is redshift dependent,
being slightly smaller than $1.686$ at low $z$ \cite{1992ApJ...386L..33L,Kitayama1996}.
Smooth dynamical DE, in general, does not change this picture significantly
\cite{Weinberg2003}.

The threshold density can also be computed for clustering DE models.
If $c_{s}$ is negligible on the scales of interest, the DM and DE
have the same peculiar velocities, which allows the use of top-hat
profiles for both of them \cite{Mota:2004pa,Abramo2007,Creminelli2010,Batista:2013oca,Pace2014a,Batista2017}.
In this case, the model is described as a system of ordinary differential
equations, which can be solved numerically up to a certain threshold,
e.g., $\delta_{m}\sim10^{6}$, which then defines $z_{c}$ and $\delta_{c}$.
For a detailed discussion about the numerical computation of $\delta_{c}$,
see \cite{Herrera2017,Pace2017,Batista:2021uhb}.

When solving for the evolution of the radial profile, we observe that
the system gets unstable when $\delta_{m}\left(r=0\right)\sim10^{4}$,
which does not allow us to define a reliable threshold for a reasonable
redshift range. This is easy to understand with the following example:
in the EdS model, starting with $\delta_{m}\left(a_{i}=0.01,r=0\right)=0.01686$,
the linear evolution indicates that $\delta_{m}\left(a=1,r=0\right)=1.686$
at $a=1$. Therefore, according to the top-hat spherical collapse
model, the nonlinearly evolved contrast will diverge at the origin.
This behavior is critical for the evolution of the whole profile,
generating spurious oscillations. 

Given this difficulty, we will use an alternative method to compute
the threshold density, which was proposed in \cite{Lee2010b} and
also developed in \cite{Batista2017} for clustering DE. Instead of
determining $\delta_{m}^{L}\left(z_{c}\right)$, we will determine
the linearly evolved contrast at $z_{\rm v}$, the redshift of virialization.
In EdS we have
\begin{equation}
\delta_{{\rm v}}\equiv\bar{\delta}^{L}{}_{m}\left(z_{\rm v}\right)\simeq1.583\,,
\end{equation}
\begin{equation}
\Delta_{{\rm v}}\equiv\frac{\rho_{m}}{\bar{\rho}_{m}}=1+\bar{\delta}^{NL}{}_{m}\left(z_{\rm v}\right)\simeq146.8\,.
\end{equation}
In the context of non-top-hat profiles, the contrasts with overbar
can be understood as volume-averaged quantities. Since at $z_{\rm v}$
the central region of the density contrast has not formally diverged,
naturally, the evolution of the entire profile does not present any
instability. 

As discussed in \cite{Batista2017}, in the presence of clustering
DE, the natural generalization for the virialization threshold is
given by
\begin{equation}
\delta_{{\rm v}}\left(z\right)\equiv\bar{\delta}_{{\rm tot}}^{L}\left(z_{v}\right)=\bar{\delta}_{m}^{L}\left(z_{{\rm v}}\right)+\frac{\Omega_{de}\left(z_{{\rm v}}\right)}{\Omega_{m}\left(z_{{\rm v}}\right)}\bar{\delta}_{de}^{L}\left(z_{{\rm v}}\right)\label{eq:definition-dv}
\end{equation}
and the the virial overdensity by
\begin{equation}
\Delta_{{\rm v}}=\Omega_{m}\left[1+\bar{\delta}_{m}^{NL}\left(z_{{\rm v}}\right)\right]+\Omega_{de}\bar{\delta}_{de}^{NL}\left(z_{{\rm v}}\right)\,.
\end{equation}
In these expressions, $z_{{\rm v}}$ is the redshift of virialization,
determined at the moment that the virial equation for non-conserving
mass is satisfied 

\begin{equation}
\frac{1}{2M_{{\rm tot}}}\frac{d^{2}M_{{\rm tot}}}{dt^{2}}+\frac{2}{M_{{\rm tot}}R}\frac{dM_{{\rm tot}}}{dt}\frac{dR}{dt}+\frac{1}{R^{2}}\left(\frac{dR}{dt}\right)^{2}+\frac{1}{R}\frac{d^{2}R}{dt^{2}}=0\,,\label{eq:basse-virialization-1}
\end{equation}
where $M_{{\rm tot}}=M_{\,m}+M_{de}$,
\begin{equation}
M_{m}=\frac{4\pi}{3}R^{3}\bar{\rho}_{m}\left(1+\bar{\delta}_{m}^{NL}\right)\,.\label{eq:mass-top-hat}
\end{equation}
and
\begin{equation}
M_{de}=\frac{4\pi}{3}R^{3}\bar{\rho}_{de}\bar{\delta}_{de}^{NL}\left(1+3c_{s}^{2}\right)\,.
\end{equation}
In the SC model, $M_{m}$ is conserved, but $M_{de}$ is not. For
more details about this implementation, see \cite{Basse2011,Batista2017}.

Now we have to define how to compute the linear and nonlinear volume
averaged contrasts, $\bar{\delta}_{m}$ and $\bar{\delta}_{de}$.
In the case of a top-hat profile, indicated by $\delta_{m}^{th}\left(t\right)$,
we have
\begin{equation}
M_{m}=\frac{4}{3}\pi R^{3}\bar{\rho}_{m}\left(1+\delta_{m}^{th}\left(t\right)\right)\,.
\end{equation}
Assuming mass conservation within the physical radius $R$, we get
the usual continuity equation
\begin{equation}
\dot{\delta}_{m}^{th}+3\left(1+\delta_{m}^{th}\right)\left(\frac{\dot{R}}{R}-\frac{\dot{a}}{a}\right)=0\,,
\end{equation}
which gives the dependency of $R$ with $\delta_{m}^{th}$ 
\begin{equation}
\frac{1+\delta_{m}^{th}\left(t\right)}{1+\delta_{m}^{th}\left(t_{i}\right)}=\left(\frac{a\left(t\right)}{a\left(t_{i}\right)}\frac{R\left(t_{i}\right)}{R\left(t\right)}\right)^{1/3}\,.
\end{equation}

For general profiles, however, we can not analytically determine the
relation between $R$ and $\delta_{m}$ because the nonlinear effects
and the presence of DE fluctuations change the profile during the
evolution. Thus we need to numerically compute the integral 
\begin{equation}
M_{m}=4\pi\bar{\rho}_{m}\int_{0}^{R}drr^{2}\left(1+\delta_{m}^{NL}\left(r,t\right)\right)\,,
\end{equation}
many times at each time of interest to determine the value of $R$
that conserves the mass. Given that the profile implemented is steep,
and that it get's much steeper in the nonlinear regime, the computation
of such integral can be numerically unstable in general. To save computational
time and for the sake of numerical stability, we determine $R$ using
\begin{equation}
M_{m}=\frac{4\pi}{3}\bar{\rho}_{m}R^{3}\left(1+\delta_{m}^{NL}\left(r_{f},t\right)\right)\,,
\end{equation}
where $r_{f}\ll\sigma$, so that the profile is nearly constant between
$0<r<r_{f}$.

With this simplification, we lose the precise association between
$M_{m}$ and the physical scale $R$, but, as we will see, the time-dependent
quantities ($\delta_{\rm v}$, $\Delta_{\rm v}$) are determined with good
accuracy. In the general case, $\delta_{\rm v}$ and $\Delta_{\rm v}$ would
also depend on the mass (or radius) scale. In the examples we will
show, we can consider that these quantities are determined for comoving
scales, $r$, such that $\delta_{m}$ is roughly constant. From figure
\ref{fig:halos_dm_de}, we can estimate
this is roughly valid for $r<0.25\text{Mpc/h}$. A more detailed analysis
of the dependence of the threshold and virialization densities on
the scale will be done in a forthcoming paper. For a study about the
scale-dependent SC quantities in the presence of linear DE perturbations,
see \cite{Basse2012}. With this setup, we can check how accurate
our model reproduces the classical SC results, see Appendix A.

\begin{figure}
\centering{}\includegraphics[scale=0.5]{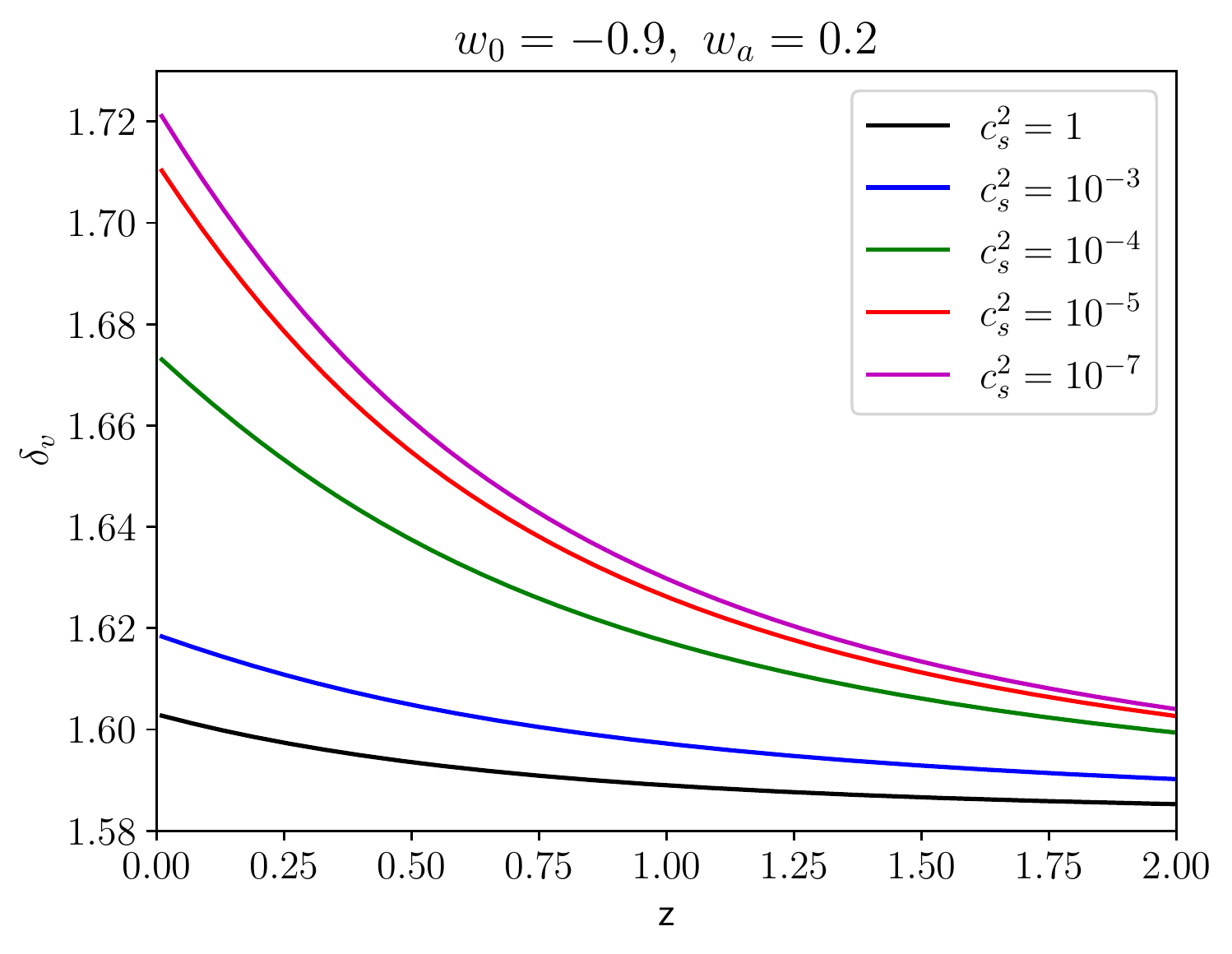}\includegraphics[scale=0.5]{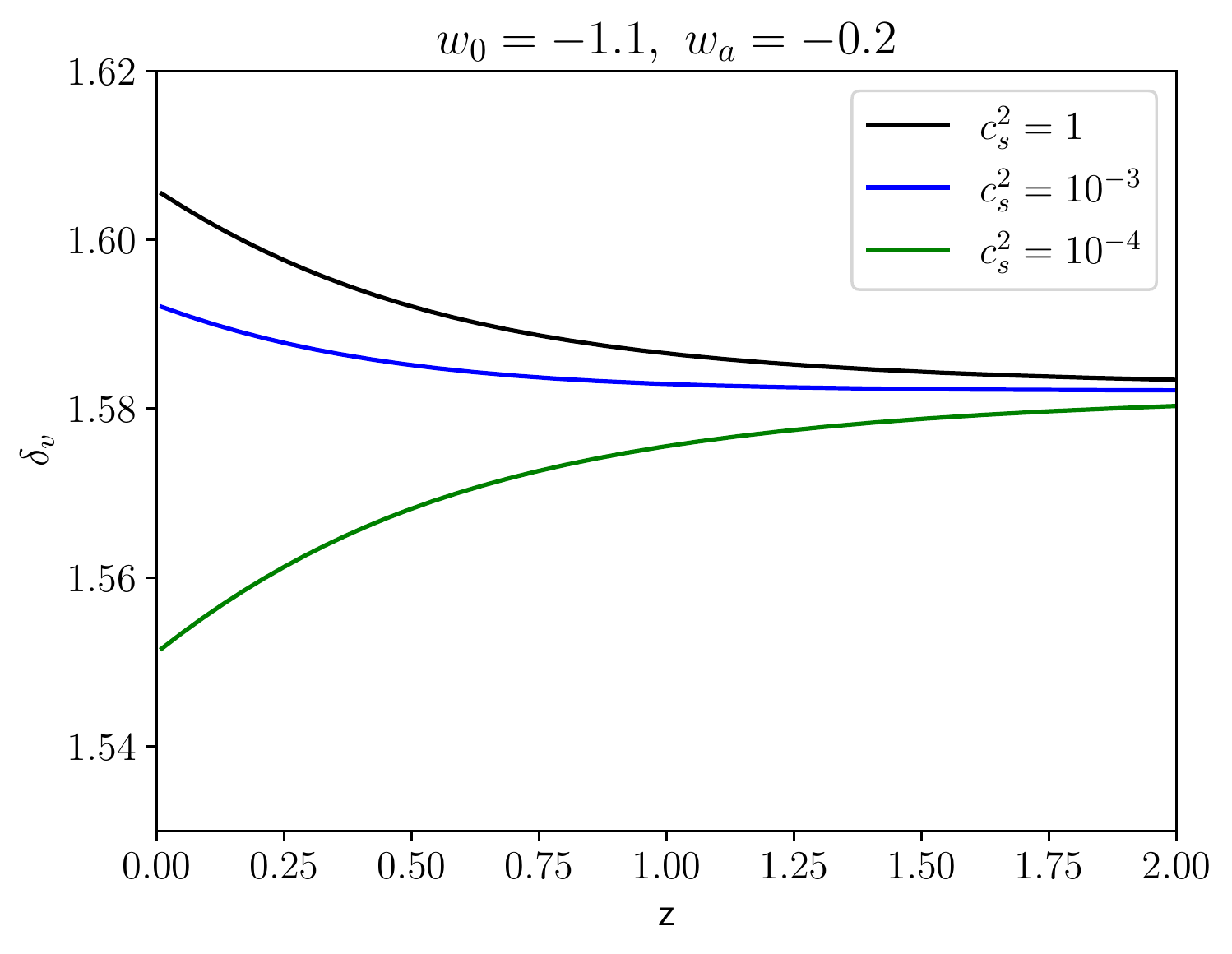}\caption{Left panel: evolution of $\delta_{{\rm v}}$ with $z$ for non-phantom
models and selected values of $c_{s}$. Right panel: the same, but
for phantom models. In this latter case, we restrict the values of
$c_{s}$, so that negative densities associated with phantom DE are
not present. \label{fig:dv-s1}\protect \\
}
\end{figure}

Finally we determine the impact of $c_{s}^{2}$ on $\delta_{\rm v}$ and
$\Delta_{\rm v}$ on small scales. We show results for a non-phantom model
($w_{0}=-0.9$ and $w_{a}=0.2$) and phantom model ($w_{0}=-1.1$
and $w_{a}=-0.2$). We verified that the values for $c_{s}^{2}<10^{-7}$
are very close with those for null sound speed. As expected, for non-phantom
DE, in both cases all curves lie in between the ones for $c_{s}^{2}=0$
and $c_{s}^{2}=1$. As can be seen in figure \ref{fig:dv-s1}, there is
an important dependence of $\delta_{{\rm v}}$ with $c_{s}$ at low-$z$.

In the phantom case, there is an interesting trend, namely, for sufficiently
small $c_{s}$, $\delta_{{\rm v}}$ decreases with $z$. This happens
because low values of $c_{s}$ will induce more negative $\delta_{de}$,
which in turn decreases the matter growth and $\delta_{{\rm v}}$.
It's also important to note that, in phantom models, DE becomes important
for the background evolution at lower redshifts. Hence, it's effects
are more apparent later than in non-phantom model. Finally, we remark
that we restricted $c_{s}^{2}\ge10^{-3}$ to avoid the negative densities
in phantom DE, as discussed in section \ref{subsec:Matter-halos}.

\section{Conclusions}

In this work, we developed a numerical code capable of solving the
nonlinear partial equations of the SC model associated with perfect
fluids with pressure. This kind of system naturally arises when some
clustering component has a scale-dependent growth, as DE with arbitrary
sound speed, and cannot be treated by the usual SC methods. Thus we
were able to generalize results in the literature that were obtained
in the limits of homogeneous DE ($c_{s}=1$) and clustering DE ($c_{s}=0$).
Our method shows very good agreement with linear solution in EdS model
and with the top-hat SC collapse model for the cases with $c_{s}=1$
and $c_{s}=0$.

We have confirmed that DE fluctuations with $c_{s}=1$ remain very
small compared to matter fluctuations, even in the nonlinear regime,
as expected in Quintessence and Tachyon models \cite{Mota:2007zn,Rajvanshi2020}.
We also verified that, for $c_{s}<10^{-5}$, DE fluctuations behave
as dust and can become nonlinear, also depending on $w$. In this
case, the evolution of matter fluctuations is strongly impacted. As
a consequence, the virialization threshold, $\delta_{{\rm v}}$, has
a substantial increase in non-phantom models and a moderate decrease
in phantom healthy models. We also found that, for $c_{s}^{2}<10^{-3}$,
the gravitational potential associated with matter halos and voids
can change about $4-9\%$ with respect to the $c_{s}^{2}=1$ case.
This can be an important observational feature of DE fluctuations
\cite{Hassani2020}. 

We have shown that phantom DE with low $c_{s}$ can develop a
pathological state of negative energy density around matter halos.
This can happen for $c_{s}^{2}<10^{-7}$ around virializations overdensities
and for $c_{s}^{2}<10^{-5}$ around overdensities $\delta_{m}\sim10^{3}$.
Therefore, in order to avoid negative densities, phantom DE models
described by perfect fluids can not have arbitraly low sound speed,
such as in model \cite{Creminelli2009}. The specific minimum value
of $c_{s}$ that avoids this pathology also depends on $w$. Thus,
helthy phantom models demand some fine tune or can not be described
by perfect fluids \cite{Sawicki2013a}. 

For the first time, we have explored the dependence of $\delta_{{\rm v}}$
with $c_{s}$. At low redshifts, the departures from the homogeneous case is about $1\%$
for $c_{s}^{2}=10^{-3}$ and increase up to $7\%$ for $c_{s}^{2}=10^{-7}$.
As shown in \cite{Batista2017}, the variation of halo abundances
between homogeneous and clustering DE can reach up to $30\%$. Therefore,
intermediate values of $c_{s}$ can also present a sizable impact
on cluster abundances. Our results are focused on small nonlinear
scales. We aim to further develop this code to precisely determine
the scale dependence of $\delta_{{\rm v}}$ and implement more realistic
profiles.

Our code can be adapted to solve the nonlinear evolution of fluctuations
in other cosmological scenarios which present scale-dependent growth
of fluctuations, such as warm DM and modified gravity models. The
inclusion of bulk viscosity is also possible and models like \cite{Velten2014a}
can be studied beyond the top-hat approximation. A particular interesting
application is the case of Ultra Light DM, \cite{Ferreira:2020fam}.
In such models, DM naturally has a scale-dependent growth, and small
halos develop a core due to ``quantum pressure''. Semi-analytical
halo abundances of these models have used prescriptions for the collapse
threshold proposed in the context homogeneus DE or warm DM \cite{Marsh:2013ywa,Marsh2016}.
Thus a more detailed semi-analytic study of the nonlinear evolution
is still lacking.

\section*{Acknowledgements}

RCB thanks João Assirati for invaluable help with the implementation
of Bash scripts and C language codes used in this project and Instituto
de Física of São Paulo University for the hospitality during the final
developments of this work. HPO acknowledges the financial support
of Brazilian Agency CNPq. LRWA acknowledges the financial support
of Brazilian Agency CNPq and São Paulo state agency FAPESP.

\section*{Appendix A: Convergence and accuracy tests}

\subsection*{Linear evolution}

\begin{figure}
\centering{}\includegraphics[scale=0.6]{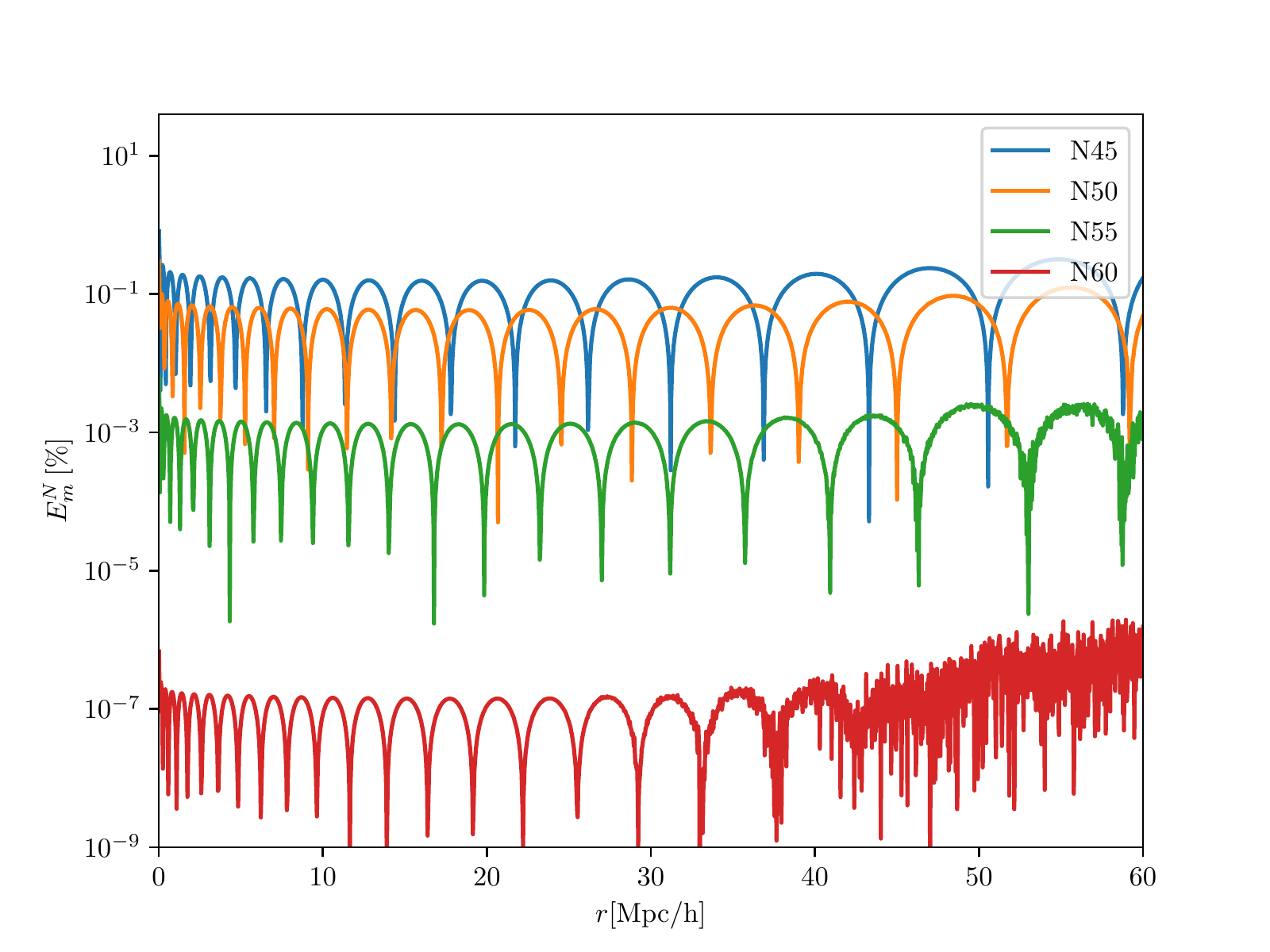}\caption{Numerical error profiles for the linear evolution $\delta_{m}$ defined
in Eq. (\ref{eq:error_mat_lin}) \label{fig:dm_error}\protect \\
}
\end{figure}

Let's first analyze the convergence and accuracy of the method in
the linear regime. In figure \ref{fig:dm_error} we show the percent
error in the $\delta_{m}$ profile at $z=0$, given by
\begin{equation}
E_{m}^{N}\left(r\right)=100\%\times\frac{|\delta_{m}^{EdS}\left(r\right)-\delta_{m}^{N}\left(r\right)|}{\delta_{m}^{EdS}\left(r\right)}\,,\label{eq:error_mat_lin}
\end{equation}
where $\delta_{m}^{EdS}$ is the analytical solution, equation (\ref{eq:mat_lin_EdS}),
and $\delta_{m}^{N}$ is the numerical solution for the truncation
order $N$. We assume that $\sigma=30\mbox{Mpc/h}$. Here we only
show the dependence with $N$, however the map parameter $L_{0}$
is also important because it changes the spatial coverage of the collocation
points, which has to be adjusted according to the profile parameters
in order to minimize the error. For the present case, $L_{0}=12$
was used.

As we can see, the error falls with the increase of the truncation
order $N$. We observe that $N=60$ minimizes the error within $1\sigma$,
which is below $10^{-6}\%$. We also note that the error is cumulative
with time. Therefore, for higher redshifts, the error is even smaller.
It is also important to note that higher values of $N$ do not necessarily
decrease the error. As $N$ grows, more numerical precision is needed
to satisfactorily evaluate the base functions at the collocation points.
In the algebraic procedure, done with Maple software, the numerical
precision can be increased to fulfil this demand. However, when exporting
the equations to be integrated with GSL routines written in C language,
we are limited to the machine precision, and the errors increase above
some $N$. The onset of this limitation can be seen as a noisy error
profile for $N=55$ and $N=60$ at large radius. For $N>60$ the error
around the origin increases, so we will use $N=60$ in the examples
shown in this paper because we mainly interest in the nonlinear evolution,
which takes place at more central regions of the profile. 

We also verified the accuracy of the numerical procedure comparing
the linearly evolved contrasts with the analytical solutions in EdS
model for DE Eqs. (\ref{eq:de_lin_IC}) and (\ref{eq:de_lin_IC_cs_Null}),
i.e., assuming that the DE is a test field in a matter-dominated universe.
For the case with $c_{s}=0$ and assuming $w=-0.9$, the error profile
is very similar to what is shown for $\delta_{m}$ in figure \ref{fig:dm_error}.

The error for $\delta_{de}$ with $c_{s}=1$ is larger, about a few
percent. Probably, this larger imprecision is due to the boundary
conditions implemented, Eq. (\ref{eq:de_boundary}), which are chosen
for better accuracy in models with low $c_{s}$, i.e, they are the
same as those for matter. For the results we have shown, the impact
of this larger error for the non-negligible sound speed cases is very
small because in this case DE perturbations are at least a few orders
of magnitude smaller than matter perturbations and barely impact its
evolution and quantities used to determine the virialization. We also
checked the error in the gravitational potential, which is about the
$10^{-5}\%$ order for $N=60$. 

\subsection*{Nonlinear evolution}

We also compute the error in the nonlinear evolution of $\delta_{m}$
as compared to the analytical solution in EdS. Let us first consider
the determination of the critical density threshold at virialization,
$\delta_{{\rm v}}$. As discussed earlier, when trying to reproduce
the usual threshold for collapse, $\delta_{c}$, severe numerical
instabilities arise. Thus, we compare the numerical and analytical
determinations of quantities at the virialization time, $z_{{\rm v}}$,
in EdS model: $\delta_{{\rm v}}\simeq1.583$ and $\Delta_{{\rm v}}\simeq146.8$.

\begin{figure}
\centering{}\includegraphics[scale=0.5]{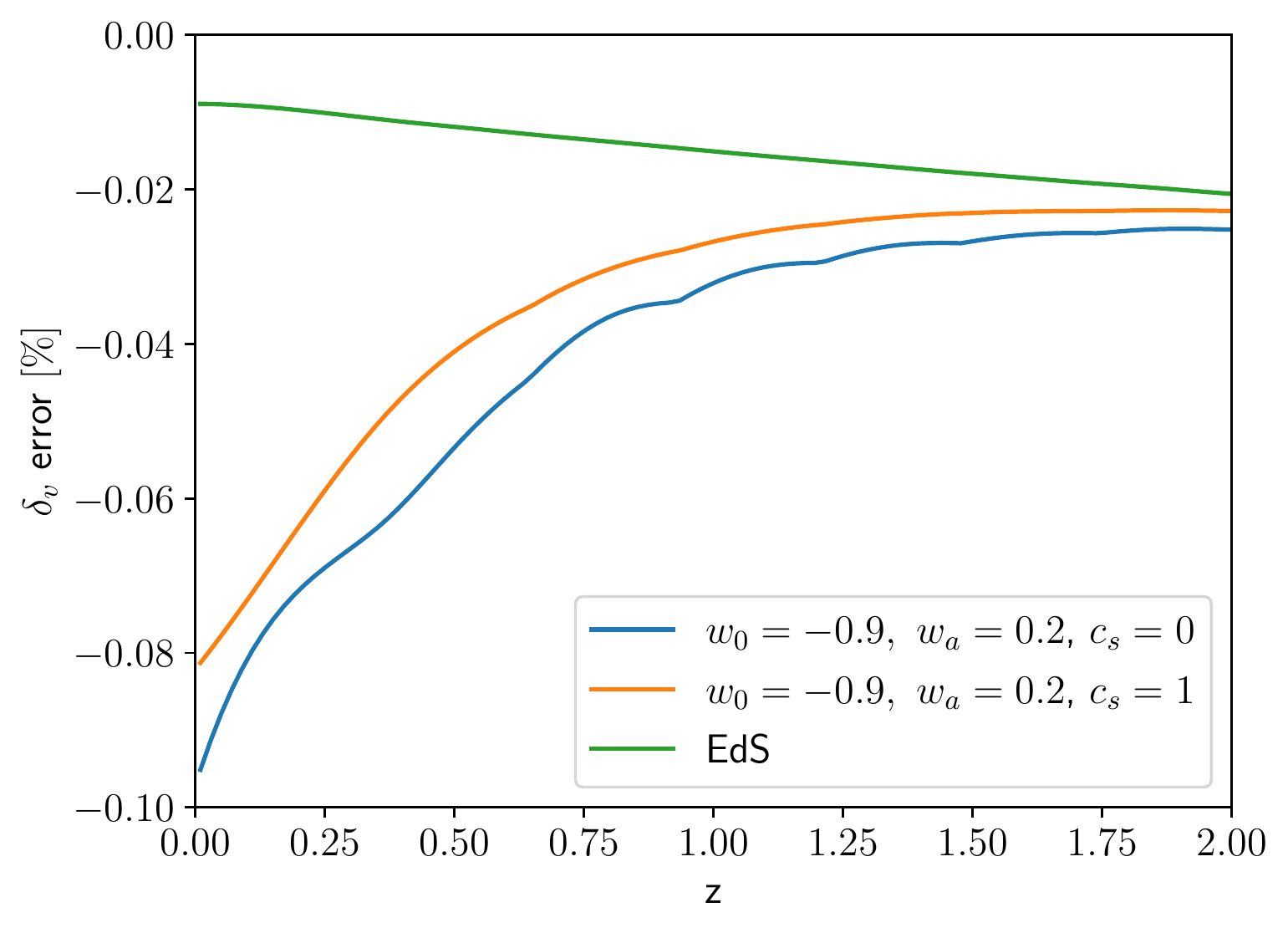}\includegraphics[scale=0.5]{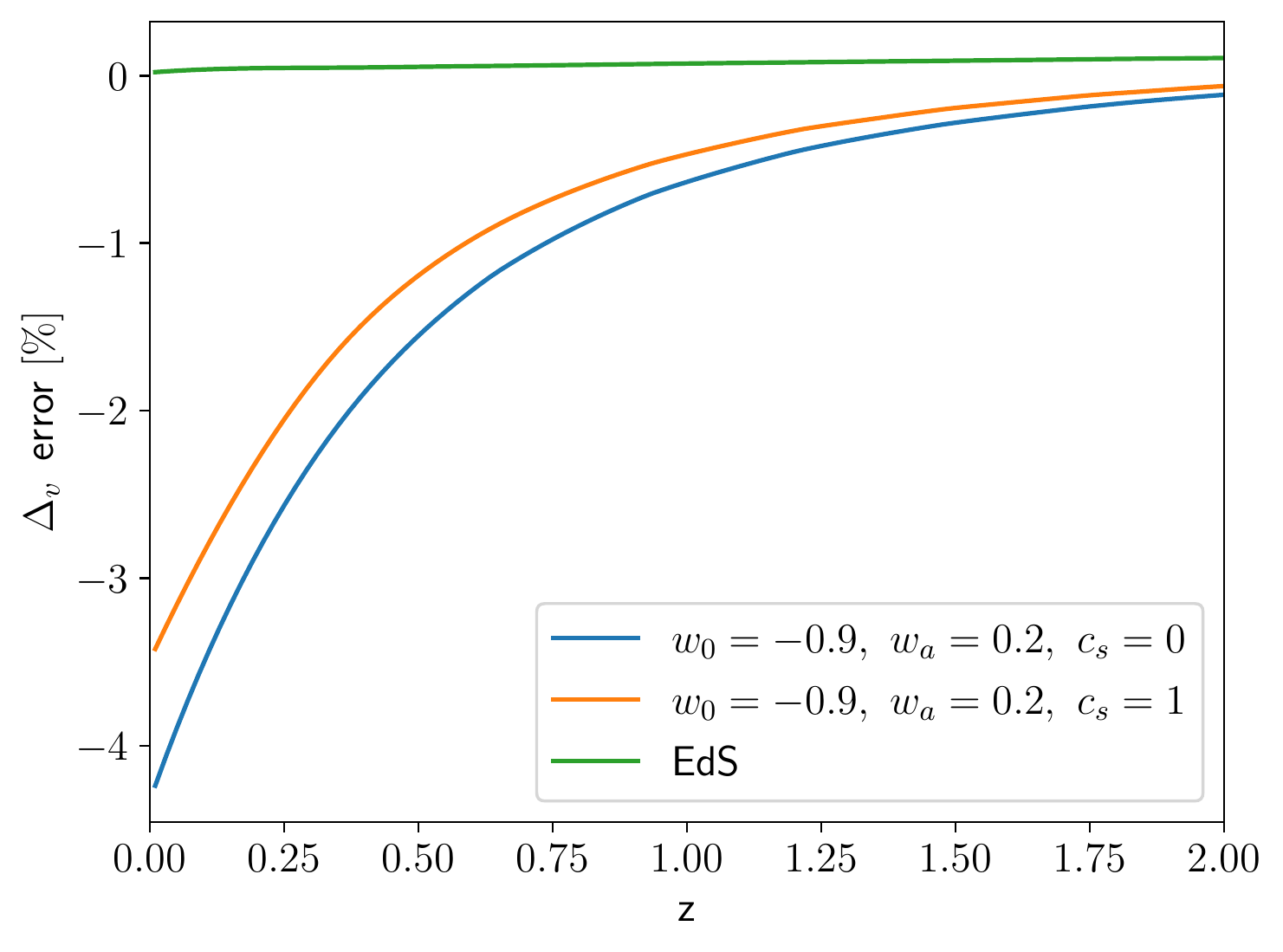}\caption{Left panel: Numerical error for $\delta_{{\rm v}}$ when compared
to the analytical solution for EdS model and the numerical solution
from the top-hat SC model for a model with $w_{0}=-0.9$ and $w_{a}=0.2$
for $c_{s}=1$ and $c_{s}=0$, as presented in \cite{Batista2017}.
Right panel: the same, but for $\Delta_{{\rm v}}$. \label{fig:dvir-error}\protect \\
}
\end{figure}

We also compare the evolution of $\delta_{{\rm v}}$ and $\Delta_{{\rm v}}$
provided by our method with the one obtained in the top-hat spherical
collapse model in the presence of clustering DE, i.e., for $c_{s}=0$
\cite{Batista2017}. In figure \ref{fig:dvir-error}, we show the
percent difference in $\delta_{{\rm v}}$ between the methods is presented
for the parameters $w_{0}=-0.9$ and $w_{a}=0.2$. As can be seen,
the errors are below $0.1\%$. For $\Delta_{{\rm v}}$ the errors
are larger, reaching a few percent for models with DE. 

The different errors magnitude for these quantities can be understood
as follows. Both of them are determined at the redshift of virialization
given by equation \ref{fig:dvir-error}. The value of $\delta_{{\rm v}}$
is then given by the linear values of the contrast, whereas $\Delta_{{\rm v}}$
by the nonlinear ones. Since $\Delta_{{\rm v}}$ is usually two orders
of magnitude larger than $\delta_{{\rm v}}$, thus the same error
in $z_{{\rm v}}$ can be amplified by roughly this amount. It's also
important to note that the errors in these quantities depend both
on the contrasts evolution and their numerical temporal derivatives,
which enter in \ref{fig:dvir-error}. Moreover, we verified that
another numerical implementation for the clustering case, based in
Python and with results shown in \cite{Batista:2021uhb}, differs
$2-3\%$ from the computations presented here and those from \cite{Batista2017},
which was implemented in Mathematica. Therefore, it's important to
note that we still lack a sub-percent accurate computation of $\Delta_{{\rm v}}$,
which can be used directly in mass functions \cite{Watson2013,Despali2016}.

\bibliographystyle{/home/rbatista/work/references/JHEP}
\bibliography{/home/rbatista/work/references/referencias}

\end{document}